\documentclass[manuscript,11pt]{aastex}
\usepackage{epsfig}

\usepackage{soul}
\usepackage{physymb}
\usepackage{mathrsfs}
\usepackage{listings}

\def\fei{Fe\,{\sc i}}

\def\fexvi{Fe\,{\sc xvi}}

\def\mathu{\textbf{\em u}}
\def\mathp{\textbf{\em p}}

\def\cm{\ifmmode {\rm cm}^{-1} \else cm$^{-1}$ \fi}
\def\s{\ifmmode {\rm s}^{-1} \else s$^{-1}$ \fi}
\def\cc{\ifmmode {\rm cm}^{-3} \else cm$^{-3}$ \fi}
\def\cs{\ifmmode {\rm cm}^{-2} \else cm$^{-2}$ \fi}
\def\g{\ifmmode \gamma \else $\gamma$\fi}
\def\G{\ifmmode \Gamma \else $\Gamma$\fi}
\def\Gs{\ifmmode \Gamma~ \else $\Gamma~$\fi}

\def\gc{\ifmmode \gamma_{\rm c} \else $\gamma_{\rm c}$ \fi}
\def\sw{Schwarzschild~}
\def\gsim{\mathrel{\raise.5ex\hbox{$>$}\mkern-14mu
             \lower0.6ex\hbox{$\sim$}}}
\def\lsim{\mathrel{\raise.3ex\hbox{$<$}\mkern-14mu
             \lower0.6ex\hbox{$\sim$}}}
\def\simless{\mathbin{\lower 3pt\hbox
     {$\rlap{\raise 5pt\hbox{$\char'074$}}\mathchar"7218$}}}   
\def\simmore{\mathbin{\lower 3pt\hbox
     {$\rlap{\raise 5pt\hbox{$\char'076$}}\mathchar"7218$}}}   
\def\Msun{M_\odot}                                
\def\4u{4U 1728--34}
\def\deg{^\circ}

\lefthead{et al.} \righthead{\4u}

\shorttitle{Fe Line Reverberation from GRHD Accretion}

\shortauthors{Porter et al.}

\begin{document}

\title{Iron Line Tomography of General Relativistic Hydrodynamic Accretion around Kerr Black Holes }

\date{\today}

\author{\textsc{Kaitlyn Porter}\altaffilmark{1,2} and 
\textsc{Keigo Fukumura}\altaffilmark{1} }

\altaffiltext{1}{Department of Physics and Astronomy, James Madison University,
Harrisonburg, VA 22807 }
\altaffiltext{2}{Email: porterkd@dukes.jmu.edu}

\begin{abstract}

\baselineskip=15pt

We consider a temporal response of relativistically broadened line spectrum of iron from black hole accretion irradiated by an X-ray echo under strong gravity. The physical condition of accreting gas is numerically calculated in the context of general relativistic hydrodynamics under steady-state, axisymmetry in Kerr geometry. With the onset of a point-like X-ray flare of a short finite duration just above the accretion surface, the gas is assumed to be ionized to produce a neutral iron fluorescent line. Using a fully relativistic ray-tracing approach, the response of line photons due to the X-ray illumination is traced as a function of time and energy for different source configurations around \sw and Kerr black holes. Our calculations show that the X-ray echo on the accretion surface clearly imprints a characteristic time-variability in the line spectral features depending on those parameters. Simulated line profiles, aimed for the future microcalorimeter missions of large collecting area such as {\it Athena}/X-IFU for typical radio-quiet Seyfert galaxies, are presented to demonstrate that state-of-the-art new observations could differentiate various source parameters by such an X-ray tomographic line reverberation.

\end{abstract}



\baselineskip=15pt

\section{Introduction}

Black hole (BH) accretion plays a fundamental role in shaping energetic radiative processes seen by X-ray spectroscopies. Among those, the iron fluorescent emission line has been extensively studied from both theoretical and observational standpoints as a valuable proxy to examine the effect of curved spacetime under strong gravity of a BH. In the context of the conventional disk-line models \citep[][]{Fabian89,Laor91,Kojima91}, the Keplerian gas orbits around a central BH forming a geometrically-thin, optically-thick structure. The disk is illuminated by an external X-ray source located somewhere near the innermost disk region. Depending on the ionization state of the irradiated gas, the energy of the line photon can be different; i.e. $6.4$ keV (for neutral), $6.7$ keV (for He-like) or $6.9$ keV (for H-like). 

The most striking effect due to strong gravity on the Fe line is its (extreme) skewness in the spectral shape; i.e. gravitational redshift, special relativistic longitudinal blueshift and classical Doppler broadening in energy. As a consequence, the iron line typically exhibits a characteristic asymmetric, double-peak feature with an extended red tail \citep[e.g.][]{Fabian89}. 
Furthermore, the red tail, produced by the most redshifted photons from the innermost disk region, is very sensitive to the radius of the innermost stable circular orbit (ISCO), which is dependent on the angular momentum of the BH (i.e. BH spin parameter) in the Shakura-Sunyaev disk \citep[][]{SS73}. Therefore, the accurate observations of the line profiles are considered to be among the most powerful means to measure BH spin \citep[][]{RF08,Fabian00,BR06,Miller07}.

In the classical framework of the disk-line scenario, as described earlier, one of the underlying assumptions is the presence of the Keplerian gas with little radial motion (i.e. $v^\phi \gg v^r \sim 0$). This assumption can be  justified when radiative cooling is so efficient in the disk that liberated gravitational potential energy of the accreting gas is quickly converted into  blackbody radiation locally \citep[e.g.][]{SS73,Novikov73,Page74,Mitsuda84}. If, on the other hand, the gas energy is not radiatively dissipated, then the energy is retained by the gas and advected into the BH \citep[i.e. advection-dominated accretion flows or ADAFs, e.g.][]{NY94,Abramowicz95} or alternatively the excess energy may be deposited into and carried away by outflows \citep[e.g.][]{BB99}. There also exist a number of variants of similar non-standard accretion models \citep[e.g. convection-dominated accretion flows or CDAFs][]{Narayan00,Quataert00}. An overall common gas property among these is depicted by (substantial) non-zero radial motion (i.e. $v^r \sim v^\phi$).

Since its first discovery with {\it ASCA} (resolving power of 120 eV; \citealt{Tanaka95}), it is observationally known that the presence of such a relativistically broadened iron line appears to be ubiquitous among various astrophysical systems hosting compact objects; i.e. Seyfert 1 AGNs \citep[e.g.][]{RF08}, galactic BH binaries \citep[][]{Miller07}, low-mass X-ray binaries (LMXBs) of neutron stars \citep[][]{Cackett08} and even white dwarfs \citep[][]{Ishida09}. 
Furthermore, owing to state-of-the-art X-ray data from the latest missions (i.e. {\it Chandra, XMM-Newton} and {\it Suzaku}) in recent years, a large number of exploratory studies of the broad line profiles have been conducted to better understand accretion physics in close proximity to the BH and its spin demographics among AGNs as well as across different mass scales \citep[][]{Nandra07}. 

In general, Seyfert 1 AGNs are observed to exhibit rapid flaring events on timescales as short as hours to days, for example, in radio-quiet Seyferts such as MCG-6-30-15 and NGC~4151, where the doubling time of the observed X-ray flux can only be thousands of seconds ($\sim 1$ ks) or even shorter \citep[e.g.][hereafter, YR00]{Lee99,YR00}.
Such a short timescale of the observed flares is comparable to the light-crossing time near the event horizon of $10^8 \Msun$ BH corresponding to the innermost accretion region. While the exact physical identity of the flares is still unknown, X-rays from such a flare are then expected to progressively irradiate the underlying accreting gas to cause fluorescence of iron. Therefore, the resulting iron line spectrum should in principle be time-dependent as the ``X-ray echo" sweeps through different parts of the gas, which would be observationally detectable if a detector is indeed capable of tracking temporal changes in line photons with a suitable timing resolution. In reality, however, low count rate of typical AGNs almost necessarily requires long exposure (e.g. at least $\sim 10$ ks or longer) to obtain statistically meaningful data with sufficiently high signal-to-noise ratios. Such a long exposure would register a series of flares all together over hundreds of light-crossing times at a gravitational radius (i.e. $t_g \equiv r_g/c = GM/c^3 \simeq 500$ s for a $10^8 \Msun$ BH). Hence, the observed line spectrum will be time-averaged and any intrinsic line variability of shorter timescales is inevitably lost. At the same time, time-averaged spectrum will conceal any important information about X-ray flares characteristic to its ``echoes" through accretion.

If, on the other hand, observations of time-resolved broad iron lines are made possible, reverberation mapping of line variability can be utilized as a powerful diagnostic tool in an attempt to observationally reveal otherwise hidden ``echoes" of X-ray flares (\citealt{BlandfordMcKee82, Stella90, R99}, hereafter, R99; \citealt{Ruszkowski00}, hereafter R00; YR00).  For example, R99 has considered such a short X-ray flare situated somewhere above the Keplerian disk and calculated a transfer function of line photons for both \sw and Kerr BHs. They demonstrated by a general relativistic (GR) ray-tracing technique that the predicted line profiles show a series of characteristic temporal behavior that largely depends on the flare position/geometry and BH spin. To follow up their implications, YR00 subsequently applied the line reverberation method to simulate time-dependent lines relevant for bright Seyfert AGNs over exposures short enough to resolve the line variability. The simulations by YR00 aim to study a realistic plausibility of such line reverberation mapping  using an anticipated detector response function for the proposed  {\it Constellation-X} mission, whose effective area was conceptualized to be about a factor of 5 larger than that of {\it XMM-Newton}/CCD detectors. The mission was designed to have sufficient sensitivity to collect enough line photons at a statistically significant level within a light-crossing time (i.e. $\sim1$ ks), as required for reverberation. Although such a high-throughput mission is unfortunately yet to be built to date, these theoretical efforts clearly point to the importance of ambitiously investigating a rapid variable nature of iron line spectrum triggered by an X-ray source, especially if indeed the line shape is responding to the ``echoes" in accretion, as commonly postulated in many theoretical calculations of broad line profiles \citep[e.g.][]{Dauser13,WF12,FukKaz07a}.   

In this paper we consider a situation where a short X-ray flare of a finite duration $\Delta T_{f}$  progressively illuminates different parts of general relativistic hydrodynamic (GRHD) non-Keplerian flows as the X-ray ``echo" propagates through the gas. 
The flare is assumed to be located at a negligible vertical height relative to the accretion surface in a way partly similar to dwarf novae \citep[e.g.][]{PattersonRaymond85,Mukai17}, accretion shocks \citep[e.g.][]{FukKaz07b,Le16,Le18} or propagation of nuclear burning on accreting neutron stars \cite[e.g.][]{Spitkovsky02}. This assumption is hence simpler than those employed in R99, R00 and YR00, but we feel that the overall systematic response of Fe K$\alpha$ lines should not differ too much for the exact source geometry.
%
Predicted iron line profiles are calculated by keeping track of null geodesics with full GR effects using a ray-tracing approach (i.e. gravitational redshift, light bending, time dilation and frame-dragging) assuming sets of fiducial source configurations around a \sw and Kerr BH for comparison. By constructing a set of time-resolved line profiles, we simulate expected line spectra using the {\it Athena}/X-IFU microcalorimeter response matrices by assuming 2-10 keV X-ray flux 
appropriate for typical bright radio-quiet Seyfert 1 galaxies. 

The structure of this paper is as follows. In \S 2 we describe our GRHD accretion model and its basic assumptions (for the gas and the characteristics of flares), which is later used to further calculate the theoretical line spectra by incorporating all the GR effects.  
The numerical results are presented in \S 3 where we show the underlying GRHD kinematics and our theoretical line profiles. The corresponding simulated observations, expected from future microcalorimeters, are also shown.  We discuss in \S 4 the plausibility of the current model as well as its observational implications that can help differentiate different models by such line reverberation analysis. 


\section{Model Description}


In this work we consider 
non-Keplerian  accretion of conserved energy $E$ and angular momentum $L$ around a Kerr BH of spin parameter $a$ under steady-state, axisymmetric assumptions in the context of GRHD. Sub-sonic gas starts accreting from rest at a large distance, while gradually  accelerating through a sonic point at $r=r_c$. The gas eventually becomes super-sonic before reaching the event horizon at $r=r_H$. By numerically solving the Bernoulli equation in GRHD framework, we calculate the gas condition including its kinematics and thermodynamic property as a function of radius for a given fluid parameter set with different BH spin values.  
Kerr spacetime metric component $g_{\mu \nu}$ is described by the Boyer-Lindquist coordinates $(t, r,\theta,\phi)$
\begin{eqnarray}
ds^2 = -\left(1-\frac{2Mr}{\Sigma}\right) dt^2 - \frac{4 M a r \sin^2 \theta}{\Sigma} dt d\phi + \frac{A \sin^2 \theta}{\Sigma} d\phi^2 + \frac{\Sigma}{\Delta} dr^2 + \Sigma d\theta^2 \ , \label{eq:metric}
\end{eqnarray}
%
where $M$ is BH mass and $a$ is its angular momentum per BH mass (i.e. spin parameter) with $\Delta \equiv r^2-2 M r+a^2$, $\Sigma \equiv r^2+a^2 \cos^2 \theta$, and $A \equiv (r^2 + a^2)^2-a^2 \Delta \sin^2 \theta$. Distance in this work is expressed by the gravitational radius $r_g$ where $r_g \equiv GM/c^2=M$ in the geometrized units ($G=c=1$). Accordingly, time is measured in units of $t_g \equiv r_g/c$. Throughout this paper we use $t_g \simeq 500$ s for a $M=10^8 \Msun$ BH mass in our calculations.

\subsection{General Relativistic Hydrodynamic Accretion}

In this model we replace the standard Keplerian accretion disk \citep[][]{SS73} by a non-Keplerian HD accretion originating from larger radius. The gas starts to plunge radially inwards with non-zero radial velocity component, which can become comparable to its toroidal component near the event horizon where $r_H \equiv r_g (1+\sqrt{1-a^2})/2$. 

Here, an ideal Boltzmann gas is considered (in the equatorial plane) whose thermal pressure $P$ obeys an ideal gas law as
\begin{eqnarray}
P  &=& \frac{k_B}{\tilde{w} m_p} \rho T \ ,
\end{eqnarray}
where $\rho, \tilde{w}$  and $m_p$ are the rest-mass density of the fluid, the mean molecular weight of the composite particles, and the mass of a particle, respectively, with Boltzmann's constant $k_B$. The equation of state for the gas is assumed to be the polytropic form as 
\begin{eqnarray}
P  &=& K \rho^\gamma = K \rho^{1+1/N} \ ,
\end{eqnarray}
where the adiabatic index $\gamma=4/3$ (equivalently polytropic index $N=3$) is adopted for  relativistic fluid. A measurement of the entropy of the gas is given by $K$, closely related to temperature $T$. The HD fluid is characterized by two conserved quantities; namely, a specific total energy $E$ and its axial angular momentum component $L$ defined by
\begin{eqnarray}
E &\equiv& -\mu u_t \ , \\
L &\equiv& \mu u_\phi \ ,
\end{eqnarray}
along a streamline where $\mu = (P+\epsilon)/\rho$ is the relativistic enthalpy of the fluid and $\epsilon = \rho + N P$ is the total energy density including the internal energy term, $NP$. Note that the four-velocity normalization requires $\mathu \cdot \mathu=-1$ where $\mathu = (u^t, u^r, u^\theta, u^\phi)$ is the four-velocity of the fluid. In this work, we assume an equatorial accretion such that $u^\theta=0$. The gas, while accelerating inwards, must be transonic; i.e. passing through a sonic point ($r=r_c$)  before entering the event horizon ($r=r_H$).

In our GRHD calculations we also define the specific angular momentum of the fluid
\begin{eqnarray}
\ell &\equiv& \frac{L}{E} = -\frac{u_\phi}{u_t} \ , 
\end{eqnarray}
that is likewise conserved along the adiabatic flow if viscous dissipation is negligible. From the four-velocity normalization one can write
\begin{eqnarray}
u_t &=& \left\{\frac{1+u_r u^r}{-(g^{tt} -2\ell g^{t \phi} +\ell^2 g^{\phi\phi})}\right\}^{1/2} \ , 
\end{eqnarray}
where $g^{\mu \nu}$ is the Kerr metric tensor components in equation~(\ref{eq:metric}). By solving  for $u^r$ with ($u_t, u_\phi$) for a given $\ell$, one can further determine the azimuthal velocity component $u^\phi$ that is related to the angular velocity of the gas as $\Omega \equiv u^\phi/u^t$. 

To better illustrate GRHD accretion solutions, we calculate physically measurable three-velocities in two different local inertial frames \citep[e.g.][]{Manmoto00};  i.e., the radial component $v^r_{\rm CRF}$ can be defined as
\begin{eqnarray}
v^r_{\rm CRF} &\equiv& \left(\frac{u_r u^r}{1+u_r u^r}\right)^{1/2} \ , 
\end{eqnarray}
in a co-rotating reference frame (CRF) where a local observer is co-rotating with the gas. On the other hand, the azimuthal component $v^\phi_{\rm LNRF}$ can be defined   as 
\begin{eqnarray}
v^\phi_{\rm LNRF} &\equiv& \frac{A}{r^2 \Delta^{1/2}} \left(\Omega-\omega \right) \ , 
\end{eqnarray}
in a locally non-rotating reference frame (LNRF) where a zero-angular-momentum-observer (ZAMO) is co-rotating with a BH relative to a distant observer with $\omega \equiv -g_{t \phi} / g_{\phi \phi}$ (i.e. frame-dragging). Note that in these reference frames it is guaranteed, by definition, that $v^r_{\rm CRF} \rightarrow 1$ and $v^\phi_{\rm LNRF} \rightarrow 0$ as $r \rightarrow r_H$.

\subsection{X-ray Sources}

A number of Seyfert 1 AGNs are known to exhibit rapid variabilities on the order of as short as hours to days in their continua \citep[e.g.][]{Lee99,Boller03,Arevalo05}. In our simulations we consider the onset of a point-like stationary flare of some finite duration $\Delta T_f$ just above the accreting gas at $(r_f,\phi_f)$ around a BH at a negligible height. The assumed X-ray flare in this work is hence somewhat similar to dwarf nova outbursts often attributed to accretion instability \citep[e.g.][]{PattersonRaymond85,Mukai17}. The wavefront of the isotropic flare progressively propagates through the equatorial gas to irradiate iron atoms for fluorescence.

\subsection{Null Geodesics}

In parallel, we numerically solve the photon geodesic equations by ray-tracing method in Kerr spacetime to calculate two separate geodesics; (i) the trajectory of echoing wavefront of the flare and (ii) the paths of line photons emitted from the gas reaching a distant observer.
The geodesic equations in Kerr geometry reads as
\begin{eqnarray}
\dot{t} &\equiv& \frac{d t}{d \lambda} = \frac{1}{\Delta} \left(r^2
+a^2 +\frac{2 a^2 M}{r} -\frac{2aM b}{r} \right) \ , \label{eq:tdot}
\\
\dot{r} &\equiv& \frac{d r}{d \lambda} = \pm
\left[\frac{-\{a^2+r(r-2M)\}\{r^3+a^2(r+2M)\}}{4M^2 a^2
r-r^3\{a^2+r(r-2M)\}\sec^2 \delta}\right]^{1/2} \ , \label{eq:rdot}
\\
\dot{\phi} &\equiv& \frac{d \phi}{d \lambda} = \frac{1}{\Delta}
\left[\frac{2aM}{r}+ \left(1-\frac{2M}{r}\right) b \right] \ ,
\label{eq:phidot}
\end{eqnarray}
where $\lambda$ is the null affine parameter, $b$ is photon's
impact parameter (closely related to angular momentum), and $\delta$ measures photon's emission angle in the rest-frame of the emitters.

As a solution to the photon geodesics, we first obtain a transfer function, $\psi_{\rm flare}(t, r,\phi)$, that describes the exact portion of the gas at ($r,\theta=\pi/2,\phi$) at time $t$ that is under illumination. Thus, $\psi_{\rm flare}(t, r,\phi)$ describes the spatial distribution of the X-ray echo 
including relativistic effects; i.e. light bending and frame-dragging. The transfer function is then convolved with the line profile calculations in the rest-frame of the HD gas, as described below in \S 2.4.   

Subsequently, we calculate another null geodesics for the line photons emitted from the gas reaching the observer. We will make full use of the elliptic integrals in this part \citep[][]{Cadez98, Fanton97} for efficient computations. Using those photons arrived at the observer, we calculate the line spectrum in the observer's frame.

\subsection{Iron Line Profiles}

While a part of the gas is uniformly irradiated during an exposure time, $t_i \le t \le t_f$, where $t_i$ and $t_f$ denote the initial and final times of the X-ray illumination at a position $(r,\phi)$ in the  accreting gas with $\Delta T_f \equiv t_f-t_i$. During that time interval, iron emission line is produced by the gas in a narrow region exposed to the X-ray illumination.

The differential (observed) line flux is given by
\begin{eqnarray}
dF_{\rm obs} (\nu_{\rm obs},t_{\rm obs}) =  g^3 ~ I_{\rm em}(\nu_{\rm em}) d\Xi \ ,
\label{eq:dFobs}
\end{eqnarray}
where $d\Xi$ is the solid angle subtended by
the line emitting region in the observer's sky (i.e. apparent accreting gas area). 
The ray-tracing calculations allow us to compute the redshift factor of photons 
\begin{eqnarray}
g(r,\phi;\theta_{\rm obs}) \equiv \frac{\nu_{\rm obs}}{\nu_{\rm gas}} \equiv \frac{(\mathp \cdot \mathu)_{\rm obs}}{(\mathp \cdot \mathu)_{\rm gas}} \ ,
\end{eqnarray}
as photons are emitted from the gas at ($r, \pi/2, \phi$) all the way out to the observer at $r=\infty$. 
Here, the four-momentum of photons $\mathp$ has been used with the gas four-velocity $\mathu$ as described in \S 2.1. 
In the local rest-frame of the gas the specific
intensity of the line emission is assumed to take the delta-function
as
\begin{eqnarray}
I_{\rm em}(\nu_{\rm em}; r,\phi) \equiv I_o(r,\phi) \delta(\nu_{\rm em}-\nu_o) = g(r,\phi;\theta_{\rm obs})
I_o(r,\phi) \delta(\nu_{\rm obs}-g \nu_o) \ , \label{eq:intensity}
\end{eqnarray}
where $\nu_{\rm em}$ is the emitted photon frequency and $\nu_o$ is the
local Fe line photon frequency. In this work, the iron is assumed  to be only weakly ionized (i.e., \fei-\fexvi) in the case of a moderately weak flare (thus equivalent to $6.4$ keV in energy). 
%
By referencing to the pre-computed transfer function $\psi$ at a given time $t$, one can determine which part of the gas in accretion yields fluorescence.    
Also, we have expressed the fluorescence intensity normalization $I_o(r,\phi)$ at a given location of $(r,\phi)$ for an optically thick accreting plasma.

%
%
Therefore, the observed line flux 
at a given observation time $t_{\rm obs}$ is found by 
\begin{eqnarray}
F_{\rm obs} (\nu_{\rm obs},t_{\rm obs})  =  \iint_{\rm flare} I_o(r,\phi) \psi_{\rm flare}(t, r,\phi) g(r,\phi;\theta_{\rm obs})^4 
\delta(\nu_{\rm obs}-g \nu_o) d\Xi \ ,
\end{eqnarray}
combined with the constructed transfer function $\psi_{\rm flare}(t, r,\phi)$. Note that the irradiated flux from the flare is calculated by GR ray-tracing approach taking into account the number of photons locally striking the gas.
Considering an observation exposure of $\Delta T_{\rm obs}$, the net observed line spectrum integrated over $\Delta T_{\rm obs}$ is calculated by
\begin{eqnarray}
F_{\rm obs}^{\rm (net)} (\nu_{\rm obs},t_{\rm obs}) = \int_t^{t+\Delta T_{\rm obs}} F_{\rm obs} (\nu_{\rm obs},t_{\rm obs}) dt  \ . \label{eq:Fobs2}
\end{eqnarray}
Hence, the observed net line spectrum above is strongly determined  by both (i) redshift map $g(r,\phi;\theta_{\rm obs})$ due to the gas kinematics and (ii) transfer function $\psi_{\rm flare}(t,r,\phi)$ of the echoing X-ray. This formalism is similarly adopted in R99, R00 and YR00. 
We assume in this work, for simplicity, that the optically thick GRHD gas in accretion is at most weakly ionized by the flare such that only neutral fluorescent photons (i.e. $6.4$ keV Fe K$\alpha$ lines) are produced.

\section{Numerical Results}

By adopting a characteristic set of HD conserved quantities, we obtain a fiducial HD solution primarily focused on its kinematics responsible for determining the redshift $g$ of the line photons as discussed in \S 2.4.

\subsection{Solutions of Non-Keplerian Hydrodynamic Flows}

We follow a similar physical prescription of GRHD accretion as discussed in the literature \citep[e.g.][]{Lu97,Lu98,FT04} by specifying a set of conserved quantities, $(E,\ell)$, to numerically solve the Bernoulli equation for the primary gas property; namely, velocity field and thermodynamic property. 
As a fiducial solution, we choose $(E,\ell)=(1.004, 3.45)$ for $a=0$ and $(E,\ell)=(1.004, 2.13)$ for $a=0.99$.  It should be reminded that accretion solutions are qualitatively insensitive to these parameter values and thus we do not conduct an exhaustive parameter search in this work. 

\begin{figure}[t]
\begin{center}$
\begin{array}{cc}
\includegraphics[trim=0in 0in 0in
0in,keepaspectratio=false,width=3in,angle=-0,clip=false]{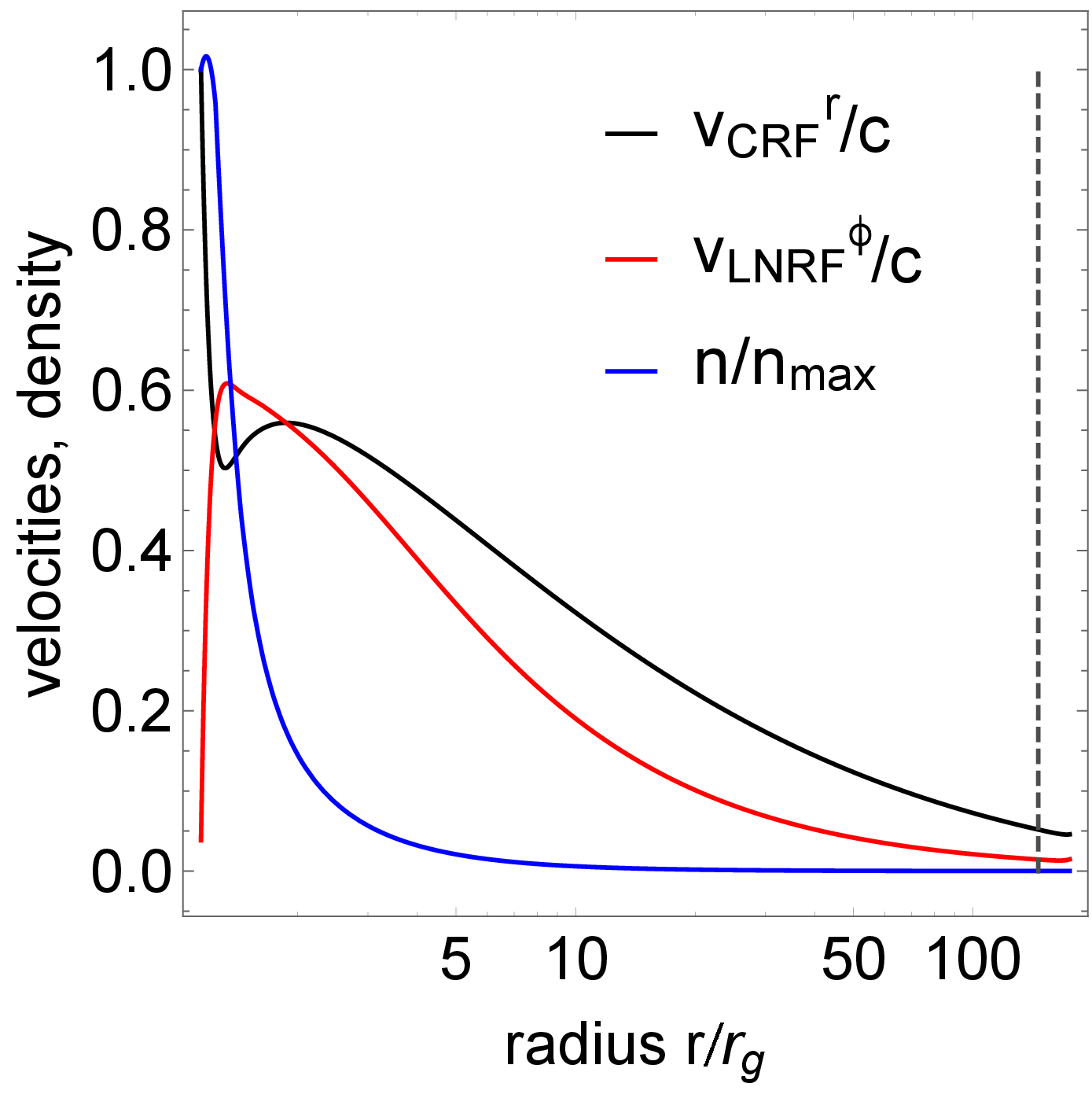}
\end{array}$
\end{center}
\caption{A fiducial solution for transonic GRHD accretion as a function of radius $r$ defined in local inertial reference frames; radial velocity $v_{\rm CRF}^r$ in CRF (black), azimuthal velocity $v_{\rm LNRF}^\phi$ in LNRF (red) and gas number density $n(r)$ (blue) normalized to its maximum value $n_{\rm max}$. We have adopted $E=1.004$ and $\ell=2.13$ for $a=0.99$ in this example. Note that $v^r \rightarrow 1$ and $v^\phi \rightarrow 0$ at $r=r_H$ as required. Vertical dotted line denotes a sonic point.  } \label{fig:accretion}
\end{figure}

As seen in {\bf Figure~\ref{fig:accretion}}, the gas slowly starts to accrete  with sub-sonic motion at a large radius where radial ($v^{\rm r}_{\rm CRF}$) and azimuthal ($v^{\rm \phi}_{\rm LNRF}$) velocity components are equally small. As it continues to fall in, the gas becomes super-sonic and both components increase as a function of radius. 
As the gas approaches the innermost region of the BH, it actually slows down due to outward external forces against the gas motion; e.g. gas pressure gradient $\grad P<0$ and the centrifugal force caused by the angular momentum barrier of the gas. In practice, a radiation pressure and magnetic force gradient would also be present (which is not included in the present calculations). Being subject to these ``obstacles", the radial acceleration decreases near the BH, but it eventually enters the horizon with the speed of light (i.e. $v_{\rm CRF}^r \rightarrow 1$ as $r \rightarrow r_H$) as demanded in the CRF \citep[e.g.][]{Abramowicz97,Manmoto00}. 
Although such a transonic flow can in principle result in shock front \citep[e.g.][]{Chakrabarti90, Chakrabarti96,Lu97, Lu98,T02,Le16}, we only consider shock-free accretion in this work for simplicity. 
In azimuthal direction, the gas orbits with its angular velocity different from that of the local spacetime. However, as it approaches the static limit ($r=2r_g$ in the equator), the frame-dragging effect inevitably guides the gas rotation such that both BH and the gas rotational velocities are getting closer to each other until they completely converge at the horizon (i.e. $v_{\rm LNRF}^\phi \rightarrow  0$ at $r \rightarrow r_H$) \citep[e.g.][]{Bardeen72,Manmoto00}, also as expected. The gas density $n$ is found to be rapidly increasing with radius $r$ (leading also to increasing gas temperature such that $\partial P / \partial r<0$).

While the quantitative behavior of the accretion solution depends on the exact values, their qualitative nature roughly remains unchanged for different values chosen. For the purpose of the current work, therefore, the choice of particular values of the fluid parameters is not important. A more detailed study of GRHD solutions are found elsewhere in the literature \citep[e.g.][]{Chakrabarti90,Chakrabarti96,Lu97,Lu98,FT04}. In \S 3.2 the line spectrum is calculated based on the calculated velocity field of GRHD accreting gas and density profile that primarily dictates the redshift distribution of the line photons and accordingly the iron line spectrum.

\subsection{Simulated Line Spectra}

In this paper we consider $M=10^8\Msun$ and $\theta_{\rm obs}=30\deg$ throughout our calculations  appropriate for archetypal radio-quiet Seyfert 1 galaxies such as MCG-6-30-15. In a simplistic prescription, we assume a point-like configuration of an X-ray flare with $\Delta T_{f}= 2 t_g \simeq 1$ ksec located at  (A) $(r_f/r_g,\phi_f)=(10,\pi)$ or (B) $(r_f/r_g,\phi_f)=(10,\pi/2)$ for a Kerr ($a=0.99$) and \sw ($a=0$) BH for comparison. We integrate the theoretical spectra to simulate a sequence of mock spectra of $2t_g \simeq 1$ks exposure over which the spectral property is not expected to significantly change. 
%
We use the publicly available response matrices for {\it Athena}/X-IFU (e.g. {\tt XIFU\_CC\_BASELINECONF\_2018\_10\_10.rmf} and {\tt XIFU\_CC\_BASELINECONF\_2018\_10\_10.arf}) to incorporate theoretical spectra into  {\tt xspec} simulations assuming a conservative X-ray flux of $f_{\rm 2-10keV} = 6 \times 10^{-12}$ erg~s~cm$^{-2}$ appropriate for typical radio-quiet Seyfert galaxies.  

\clearpage

\begin{figure}[th]
\begin{center}$
\begin{array}{cc}
\includegraphics[trim=0in 0in 0in
0in,keepaspectratio=false,width=2.5in,angle=-0,clip=false]{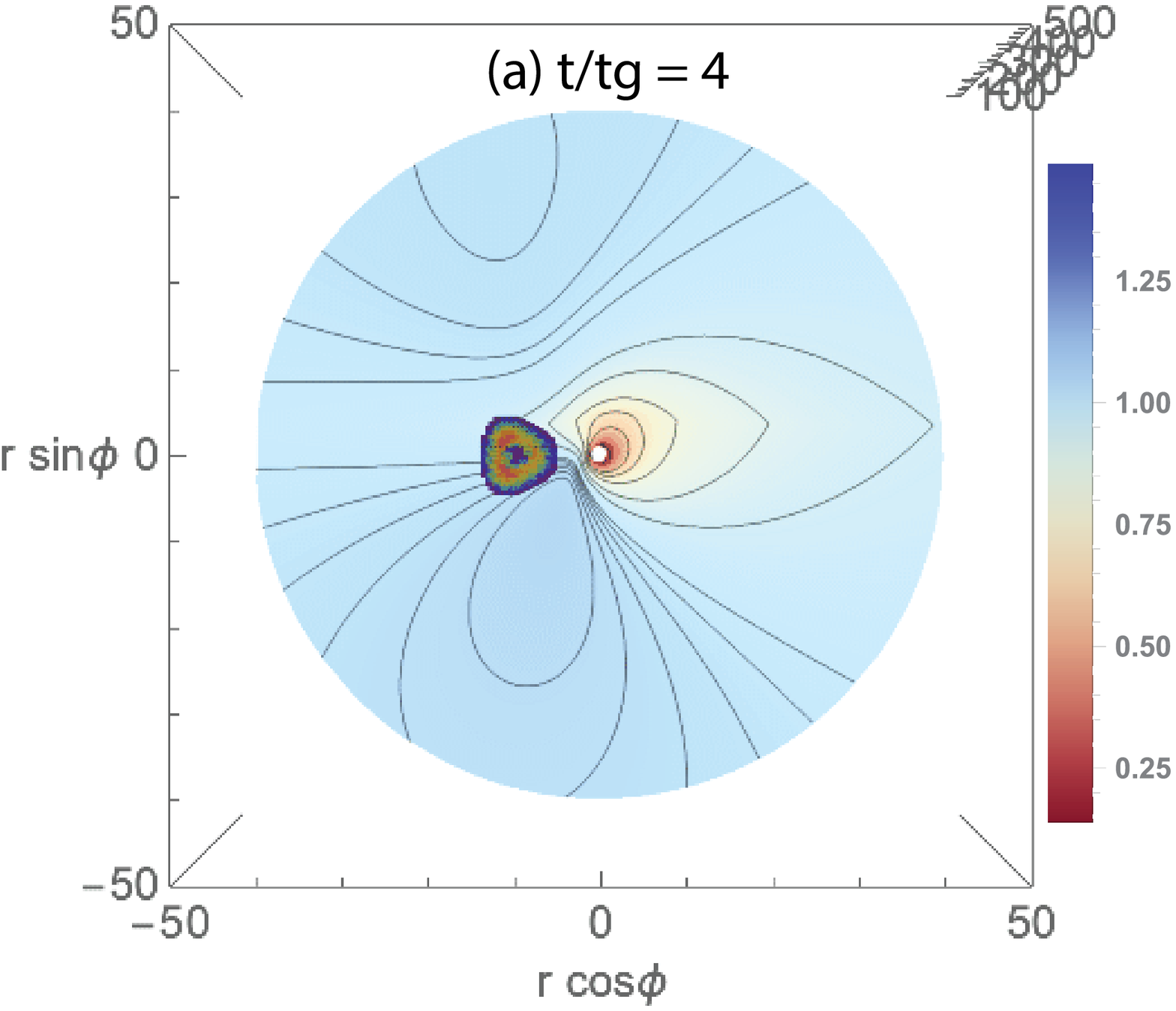}
\includegraphics[trim=0in 0in 0in
0in,keepaspectratio=false,width=2.5in,angle=-0,clip=false]{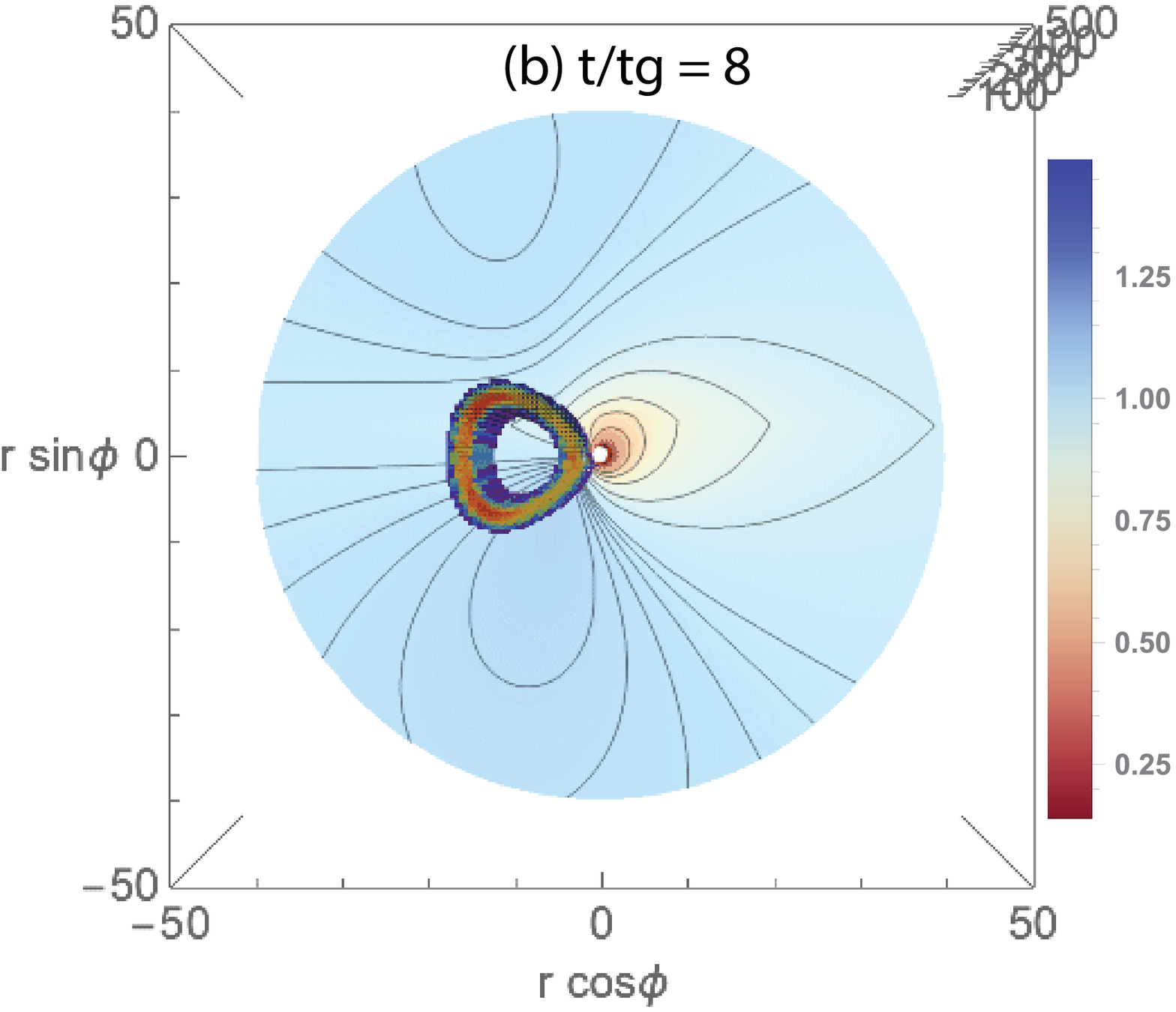} \\
\includegraphics[trim=0in 0in 0in
0in,keepaspectratio=false,width=2.5in,angle=-0,clip=false]{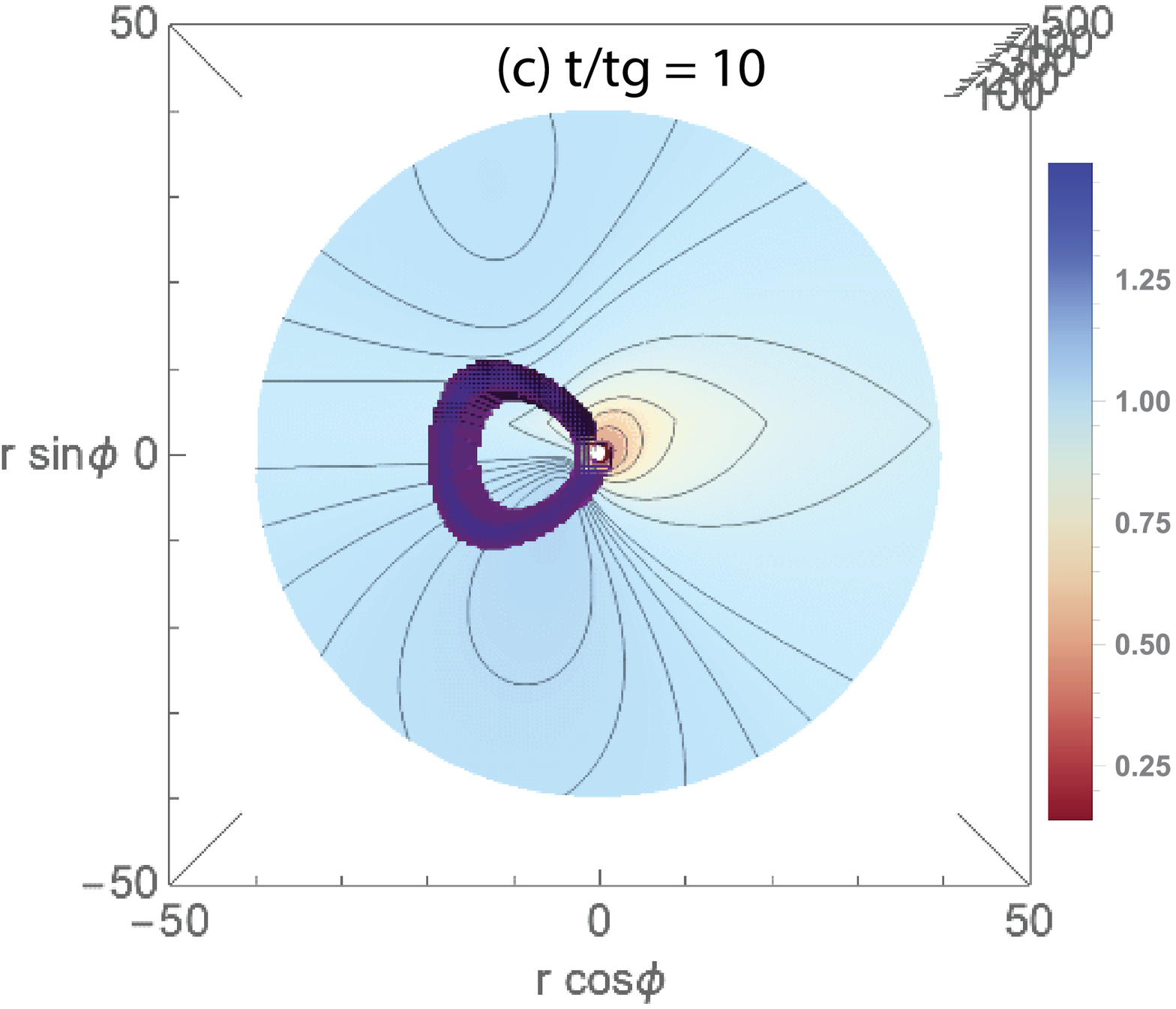}\includegraphics[trim=0in 0in 0in
0in,keepaspectratio=false,width=2.5in,angle=-0,clip=false]{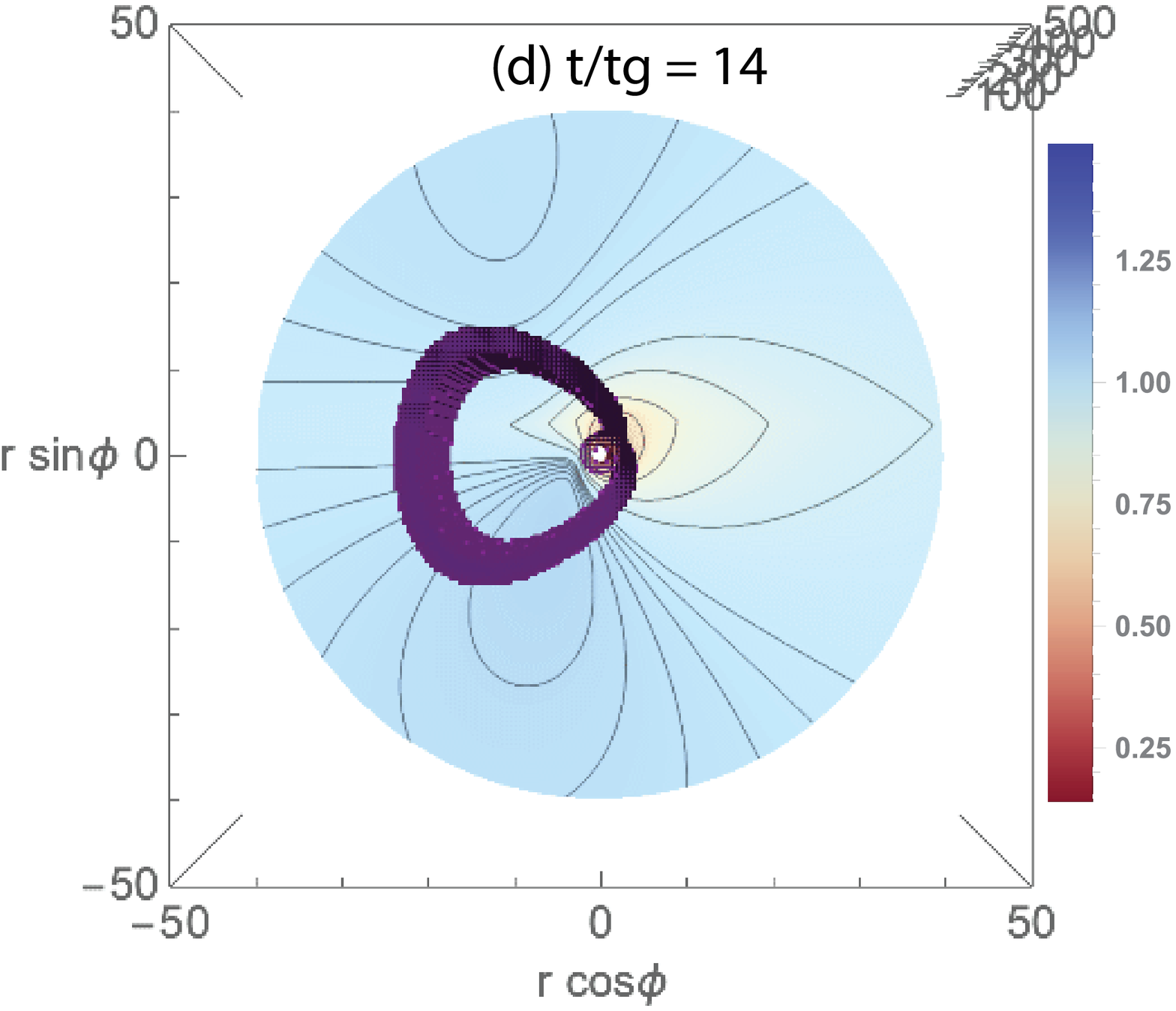} \\
\includegraphics[trim=0in 0in 0in
0in,keepaspectratio=false,width=2.5in,angle=-0,clip=false]{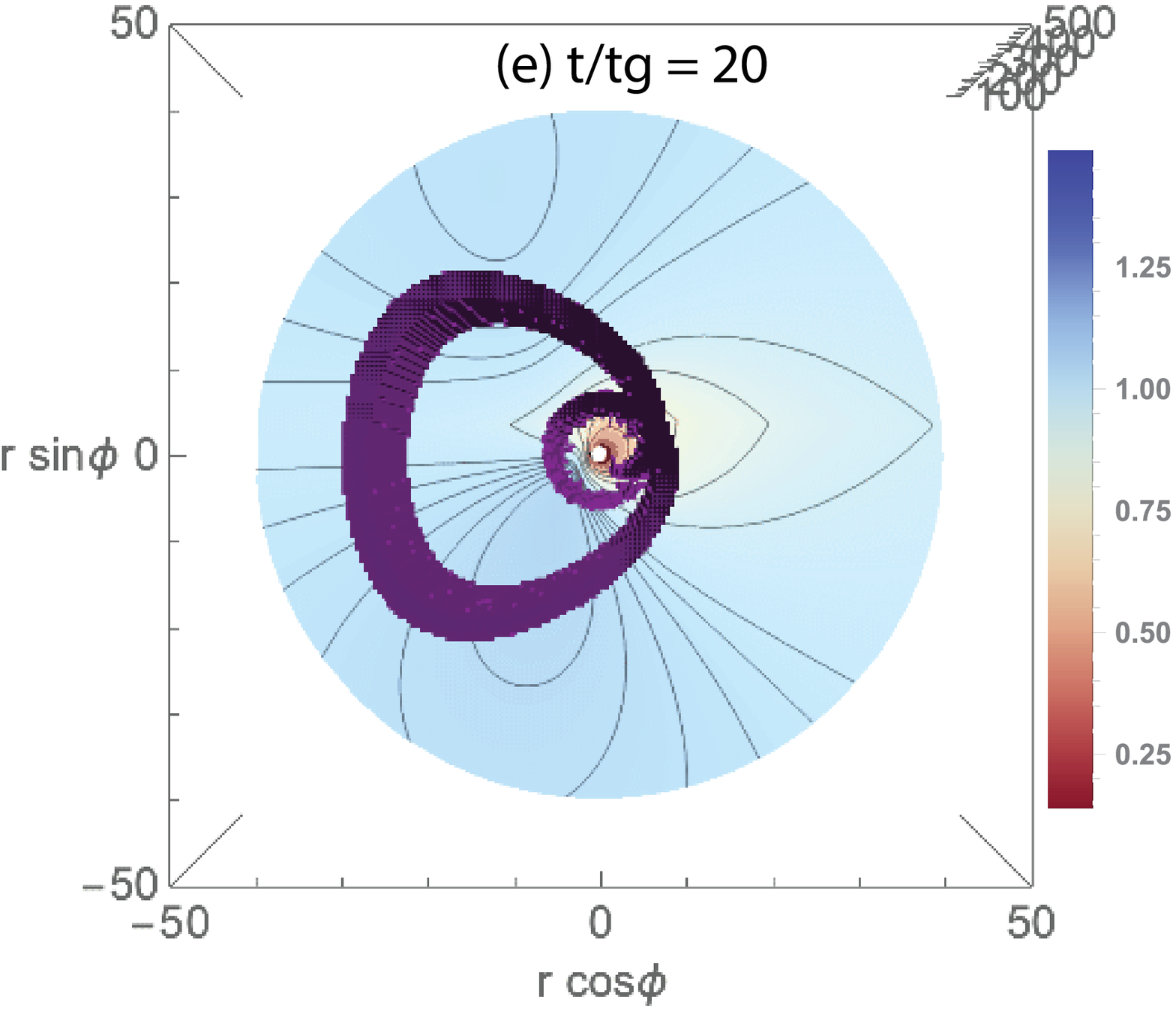}\includegraphics[trim=0in 0in 0in
0in,keepaspectratio=false,width=2.5in,angle=-0,clip=false]{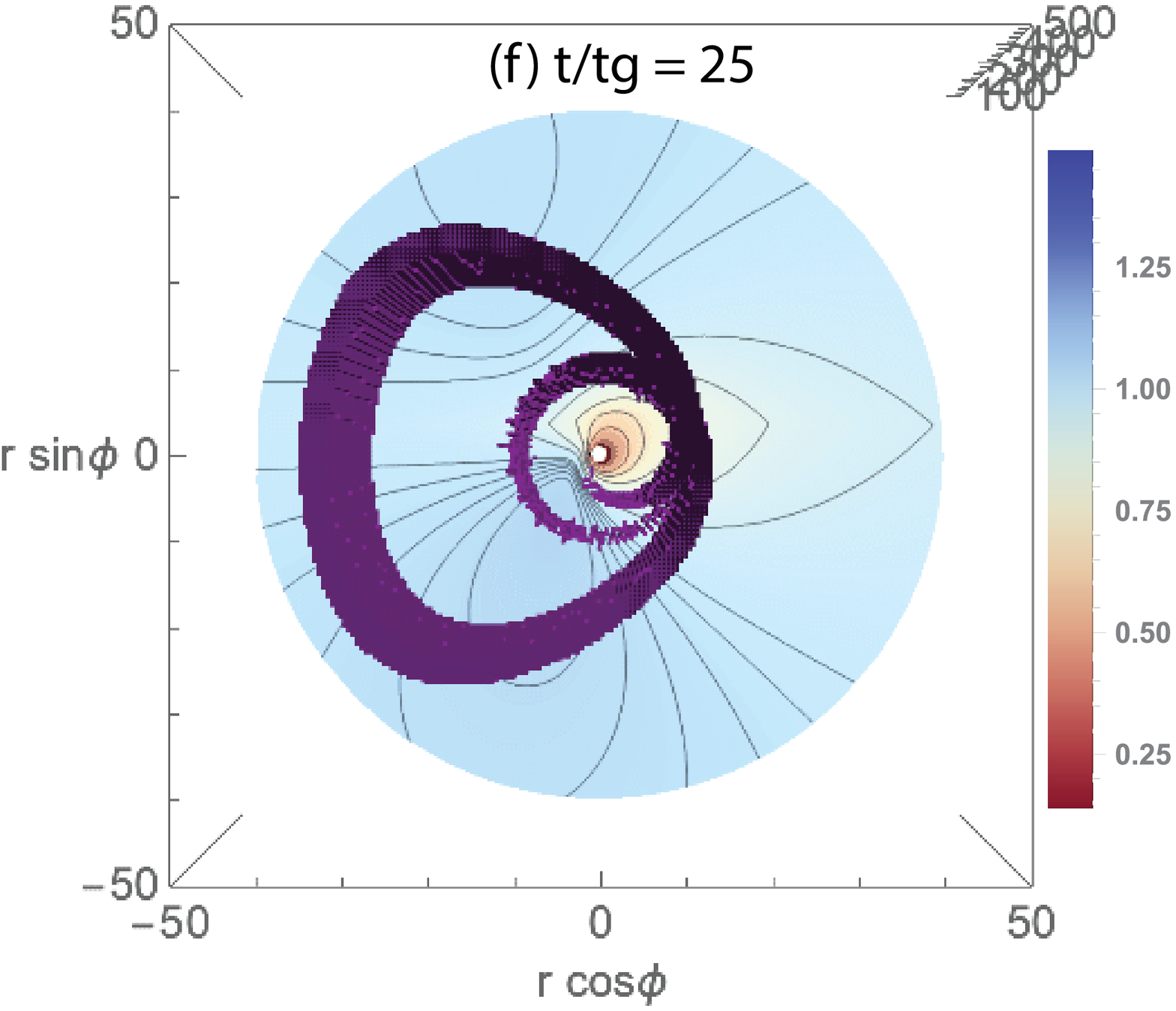}
\end{array}$
\end{center}
\caption{Redshift  maps seen with $\theta_{\rm obs}=30\deg$ for an X-ray flare of duration $\Delta T_f=1$ks in case (A) where emitted from $(r/r_g, \phi) = (10, \pi)$ echoed through accreting gas  around a Kerr BH of $a=0.99$ at $(0,0)$. Time stamp is labeled. Gas accretes counterclockwise in these face-on view maps showing more redshift of the emitted line photons in $\phi \sim 0\deg$. Color of the flare depicts the incident X-ray flux locally impinging upon the gas. 
} \label{fig:redshift}
\end{figure}

\clearpage

By numerically integrating the equatorial geodesic equations~(10)-(12), each photon emitted from the flare is  tagged by its time and position $(t, r, \phi)$. Fluorescence occurs in the irradiated part of the accreting gas, and emitted fluorescent line photons (assuming cold iron) are accordingly redshifted in energy as calculated from equation~(14). 
{\bf Figure~2} shows a sequence of the X-ray wavefront (i.e. echo) originating from an instantaneous X-ray flare at a given time, superimposed by the calculated redshift distribution from GRHD accreting gas (orbiting counterclockwise in {\bf Figure~\ref{fig:redshift}}) for case (A) also with contours in the computational domain ($0 \le r/r_g \le 50$). We assume an extreme Kerr BH with $a=0.99$ forming the static limit at $r/r_g=2$
in the equator (i.e. $\theta=\pi/2$) and a distant observer in these calculations is located at $(r_{\rm obs}/r_g,\theta_{\rm obs},\phi_{\rm obs}) = (1000, 30\deg,-\pi/2)$. 
Note that GRHD flows extend all the way down to the horizon.
%

With the onset of a flare of finite duration $\Delta T_f \simeq 1$ ks from a point-like source at $(r_f/r_g,\phi_f)=(10,\pi)$, the wavefront gradually transforms into a ``concentric ring" whose shape is subject to curved spacetime around a Kerr BH. 
Due to a finite duration of the short flare, the flare's wavefront transforms into a ``concentric ring" whose shape is constantly relativistically skewed over time. 
%
Ray-tracing indeed clearly shows that the ``ring-like" wavefront of the flare is progressively propagating through the gas in time, but the initially annular wave front is later distorted especially near the horizon due to inevitable light bending and frame-dragging, as expected (e.g. R99; R00; YR00).
%
It is seen in our calculations that the non-Keplerian GRHD accretion considered here is slightly redshifted in the most of the region. 
In fact, some regions are even more redshifted compared to those around the Keplerian accretion because the gas in our current GRHD accretion is not bound to the classical innermost stable circular orbit (i.e. ISCO at $r/r_g \sim 1$ for $a=0.99$; See \citealt{HawleyKrolik01} and \citealt{Beckwith08} for detailed global simulations on the ISCO).  
After the wavefront passes through the horizon (for example, at $t/t_g \sim 10-14$), some of the fluorescence photons are gravitationally bent and wrapped around the BH due to strong gravity and frame-dragging, breaking axisymmetry, while producing a spiral-like sub-waveform within the primary one (see at $t/t_g \sim 20-25$). In the presence of the frame-dragging effect, therefore, the innermost part of the accretion can be more effectively irradiated by the impinging flare in a concentric fashion around the Kerr BH, helping to produce more relativistically redshifted photons.



\begin{figure}[ht]
\begin{center}$
\begin{array}{cc}
\includegraphics[trim=0in 0in 0in
0in,keepaspectratio=false,width=3.0in,angle=-0,clip=false]{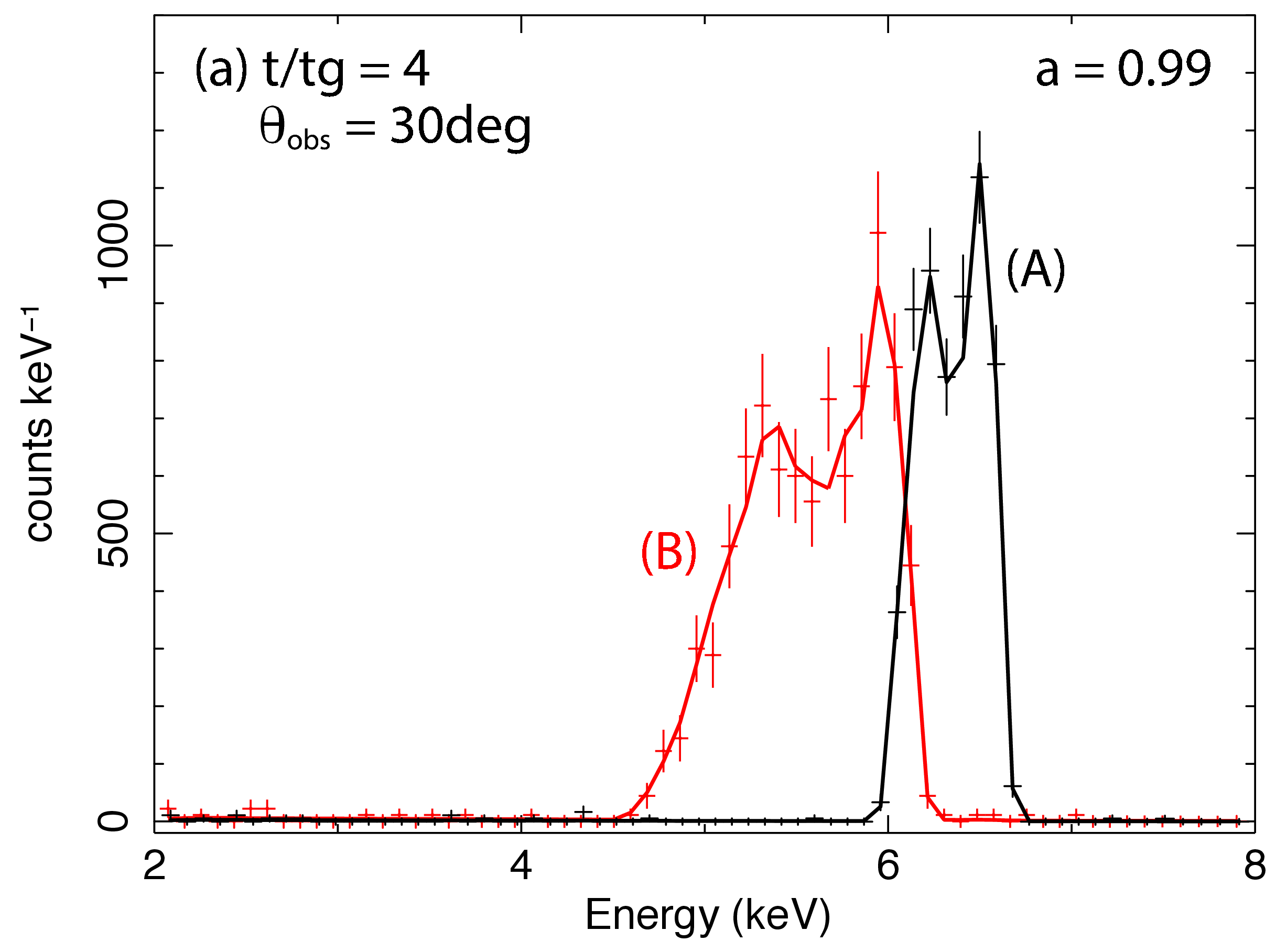}
\includegraphics[trim=0in 0in 0in
0in,keepaspectratio=false,width=3.0in,angle=-0,clip=false]{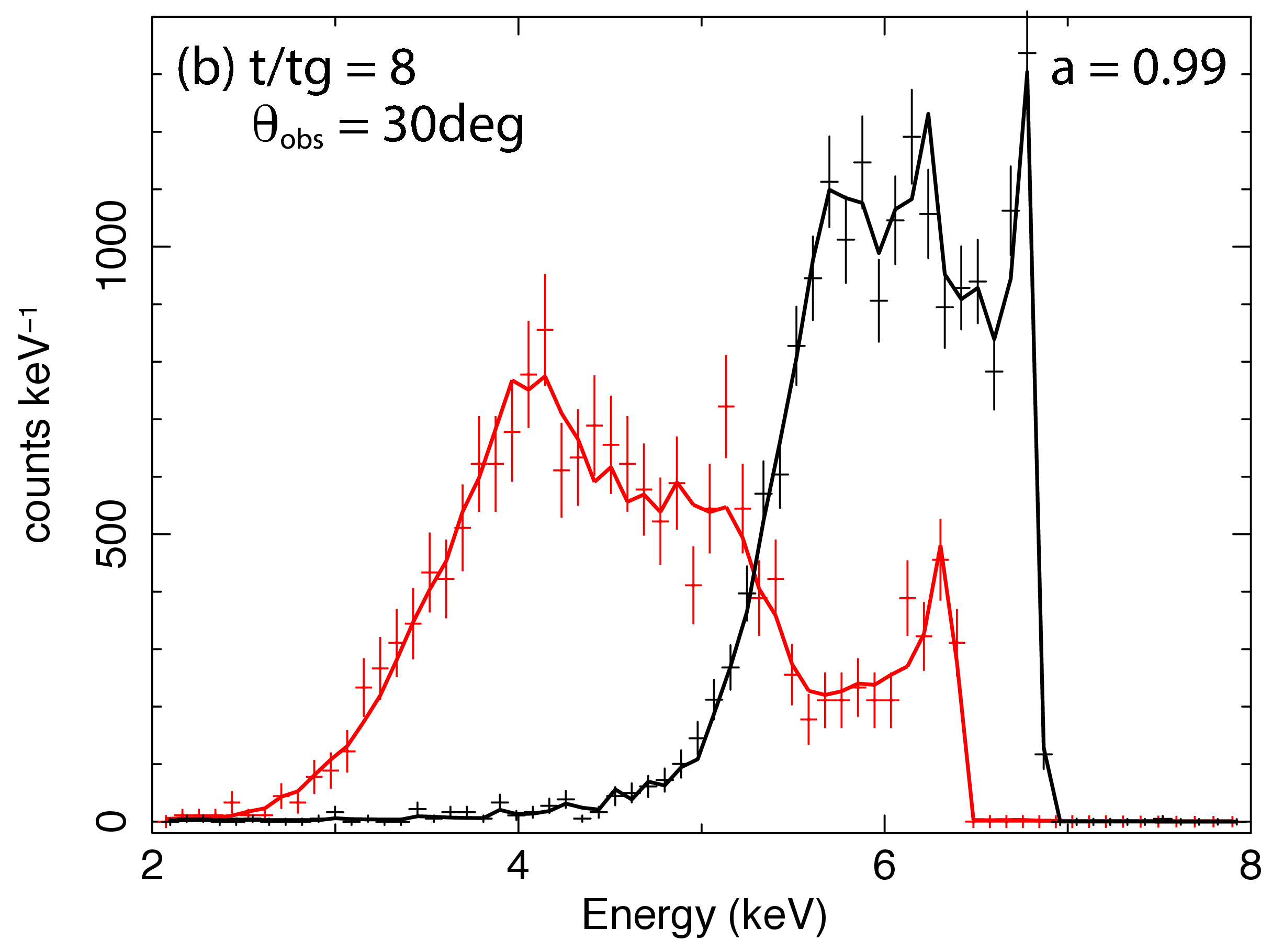} \\
\includegraphics[trim=0in 0in 0in
0in,keepaspectratio=false,width=3.0in,angle=-0,clip=false]{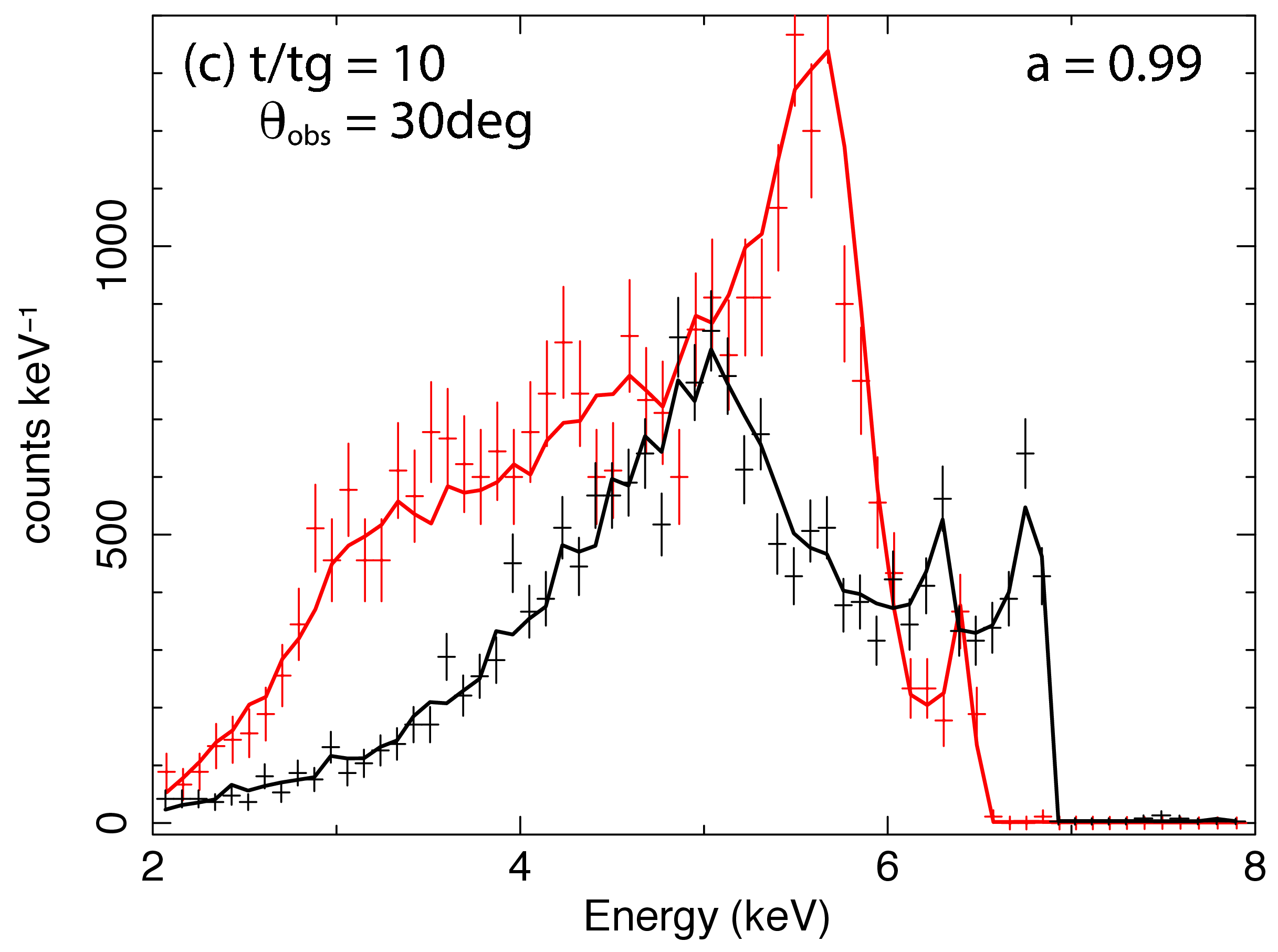}\includegraphics[trim=0in 0in 0in
0in,keepaspectratio=false,width=3.0in,angle=-0,clip=false]{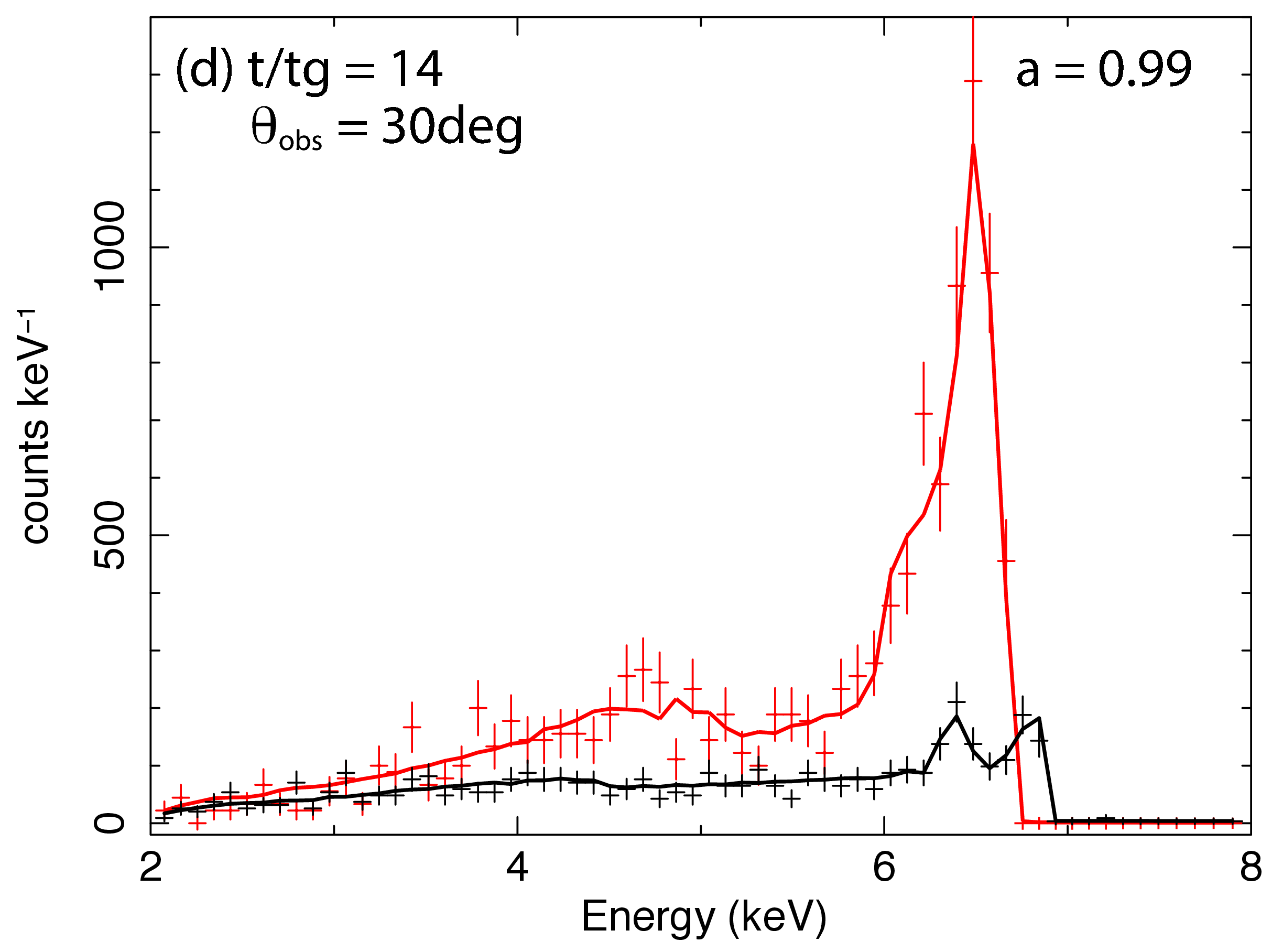} \\
\includegraphics[trim=0in 0in 0in
0in,keepaspectratio=false,width=3.0in,angle=-0,clip=false]{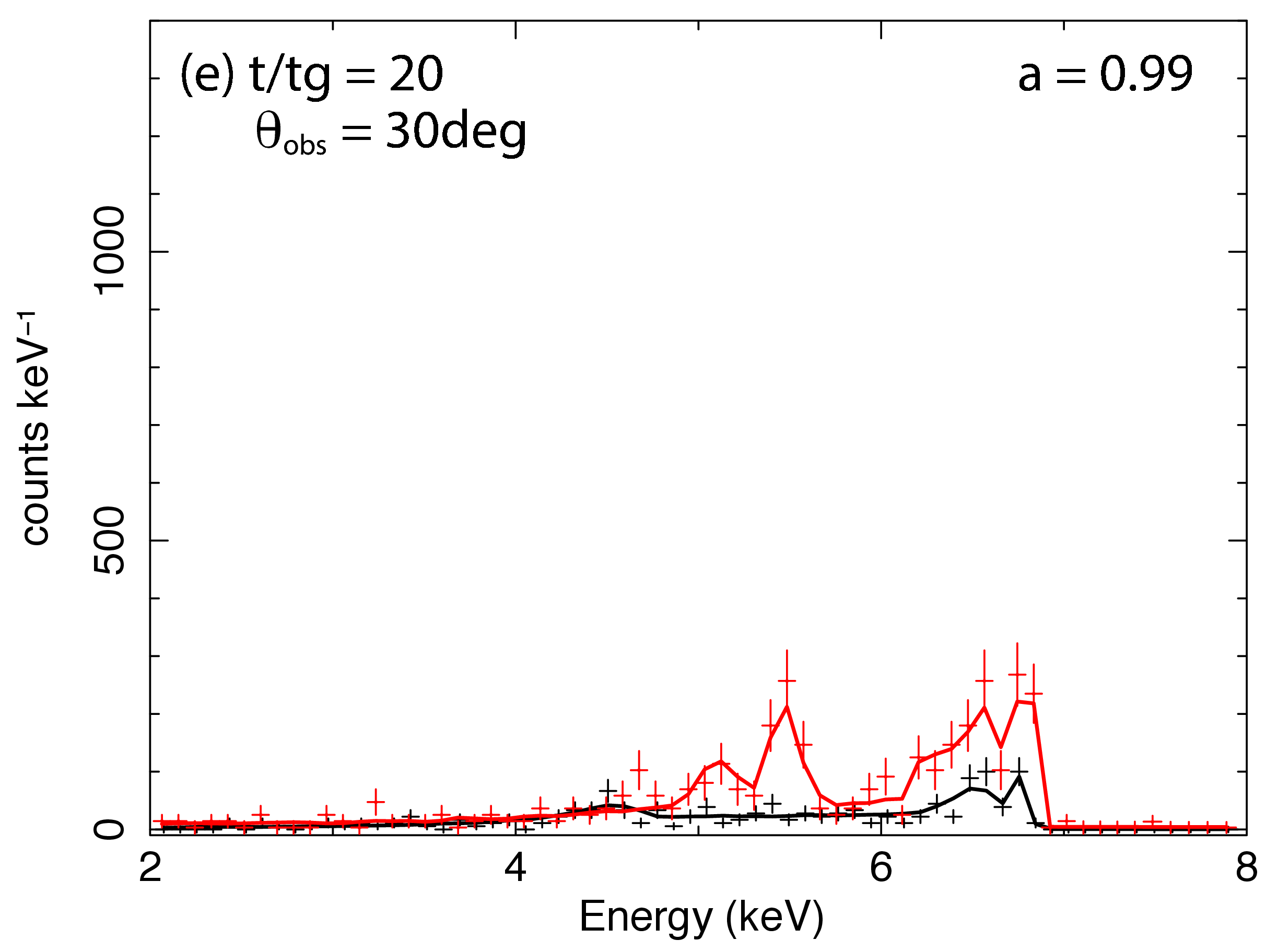}\includegraphics[trim=0in 0in 0in
0in,keepaspectratio=false,width=3.0in,angle=-0,clip=false]{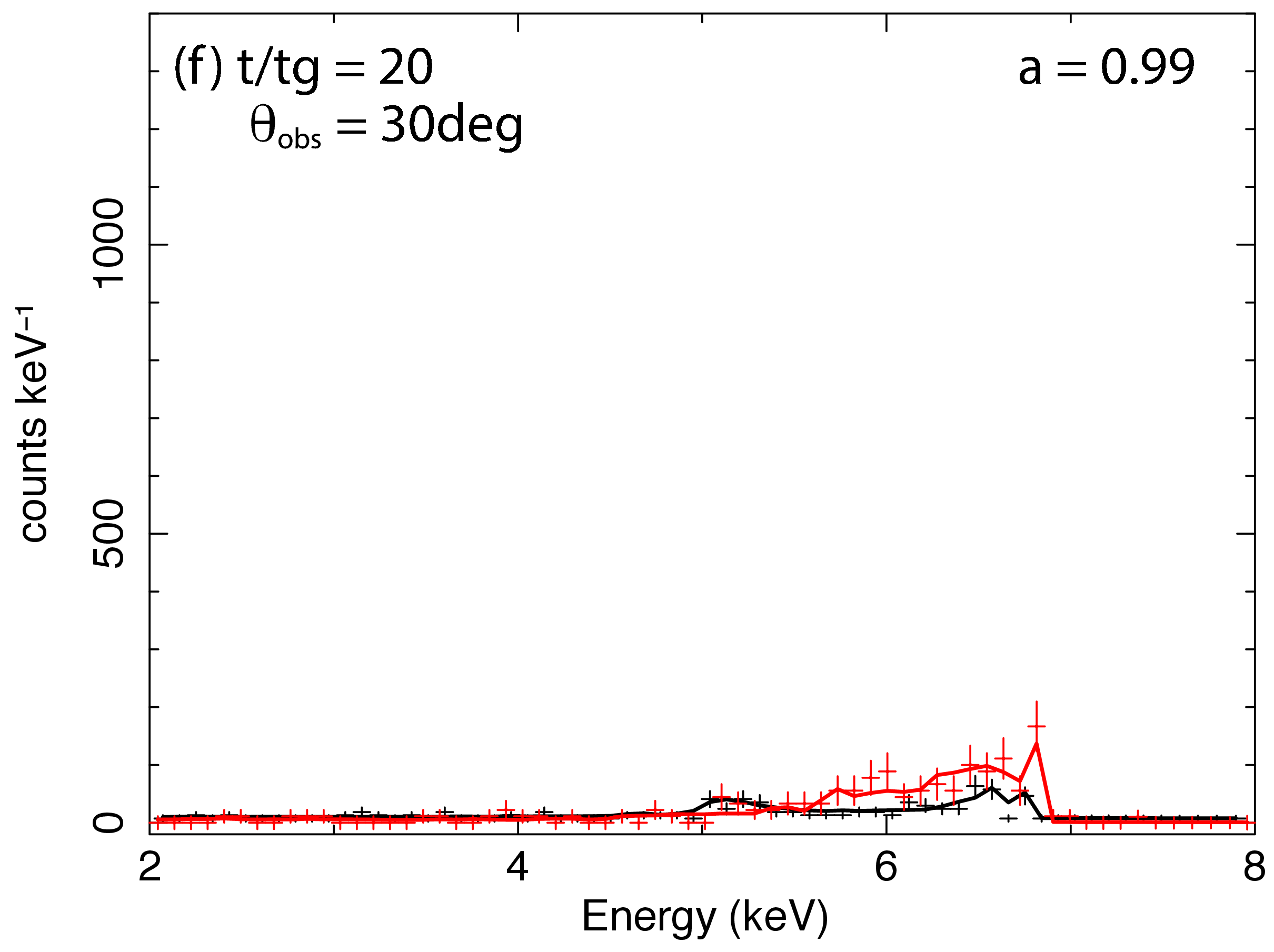}
\end{array}$
\end{center}
\caption{$1$ks simulated {\it Athena}/X-IFU spectra for $\phi_f=\pi$ (case A; black) and $\phi_f = \pi/2$ (case B; red) around a Kerr BH for the GRHD flows (solid curves) with $\theta_{\rm obs}=30\deg$. The line flux in case (A) is reduced by factor of two for presentation purpose. } \label{fig:spec_a099}
\end{figure}

\clearpage

In {\bf Figure~\ref{fig:spec_a099}} we show the calculated line profiles for $a=0.99$ and $\theta_{\rm obs}=30\deg$ for case (A) and (B). The line flux in case (A) is reduced by factor of two for presentation purpose.
The line profile initially begins  as a single narrow peak around 6.4 keV (assuming neutral Fe) with the onset of a rapid flare that acts like a point-like source. 
At this point around $t/t_g \sim 4$, the spectrum starts showing two peaks in both cases (A) and (B); the higher energy peak originates from the outer part of accreting gas under irradiation, while the lower energy peak is a result of the inner part of the gas subject to a more gravitational redshift around the BH. As seen, the spectrum is systematically less redshifted in case (A) because the wavefront is preferentially propagating through the approaching side of the accretion in the early times.  
For $4 \lesssim t/t_g \lesssim 8$, the wavefront is sweeping towards the horizon. The position of the initial red peak continuously shifts towards lower energy because of the gravitational redshift at small radii. 
After $t/t_g \sim 8$, the spectrum becomes broadened relativistically as the fluorescence occurs in the most gravitationally redshifted part of the inner accretion. It is also seen that the red peak gets  stronger relative to the blue peak later (i.e. $t/t_g \sim 8$ in red and $\sim 10$ in dark) in the  innermost region because the gas density there is higher (see {\bf Fig.~1}). As a part of the  wavefront progressively reaches the approaching side of the gas, the blue peak is accordingly enhanced due to beaming effect ($\propto g^4$) depending on exactly when this region is illuminated by the flare (e.g. $t/t_g \sim 10$ in red while $\sim 14$ in dark). 
At later times when the flare has already passed through the horizon and further continues to propagate radially outward away from the BH (e.g. $t/t_g \gsim 14$), the red peak begins to shift towards $6.4$ keV, while its flux also decreases due to decreasing gas density with increasing radius (see {\bf Fig.~1}). Although weak and subtle, this unique characteristic signature can be traced as a proxy to identify the existence of such an X-ray flaring echo. A similar trend is found and discussed in R99, R00 and YR00.

\begin{figure}[ht]
\begin{center}$
\begin{array}{cl}
\includegraphics[trim=0in 0in 0in
0in,keepaspectratio=false,width=3.0in,angle=-0,clip=false]{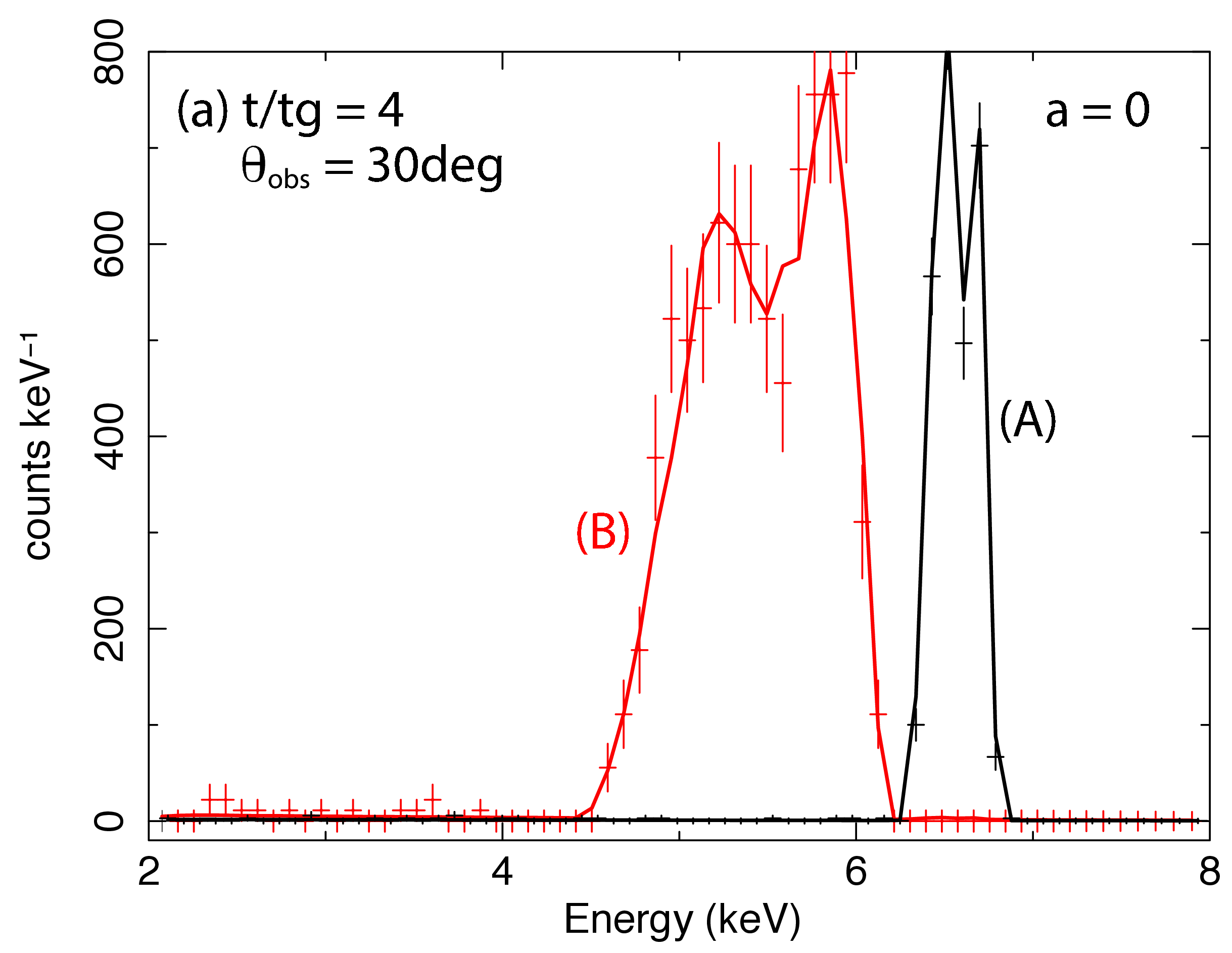}\includegraphics[trim=0in 0in 0in
0in,keepaspectratio=false,width=3.0in,angle=-0,clip=false]{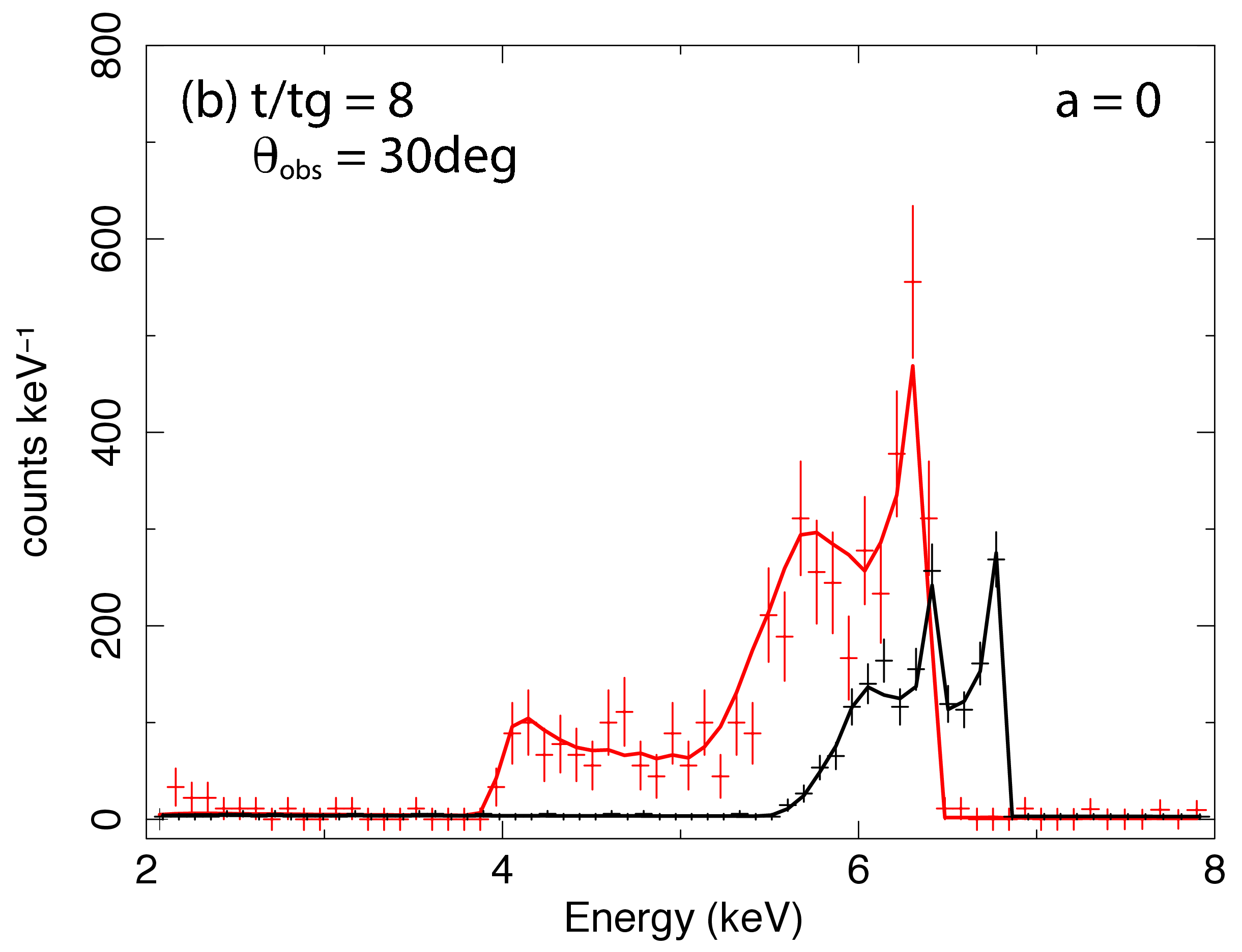} \\
\includegraphics[trim=0in 0in 0in
0in,keepaspectratio=false,width=3.0in,angle=-0,clip=false]{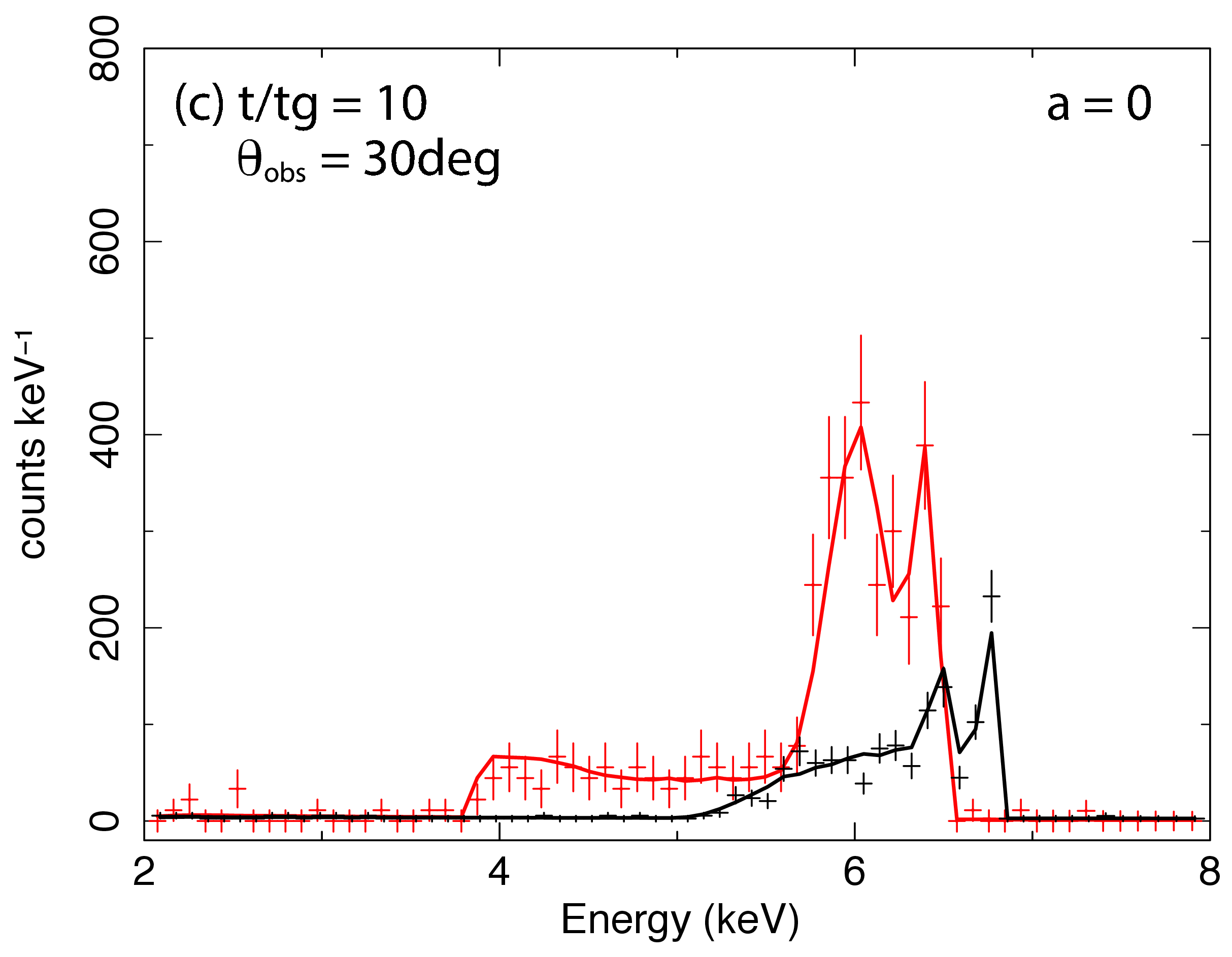}\includegraphics[trim=0in 0in 0in
0in,keepaspectratio=false,width=3.0in,angle=-0,clip=false]{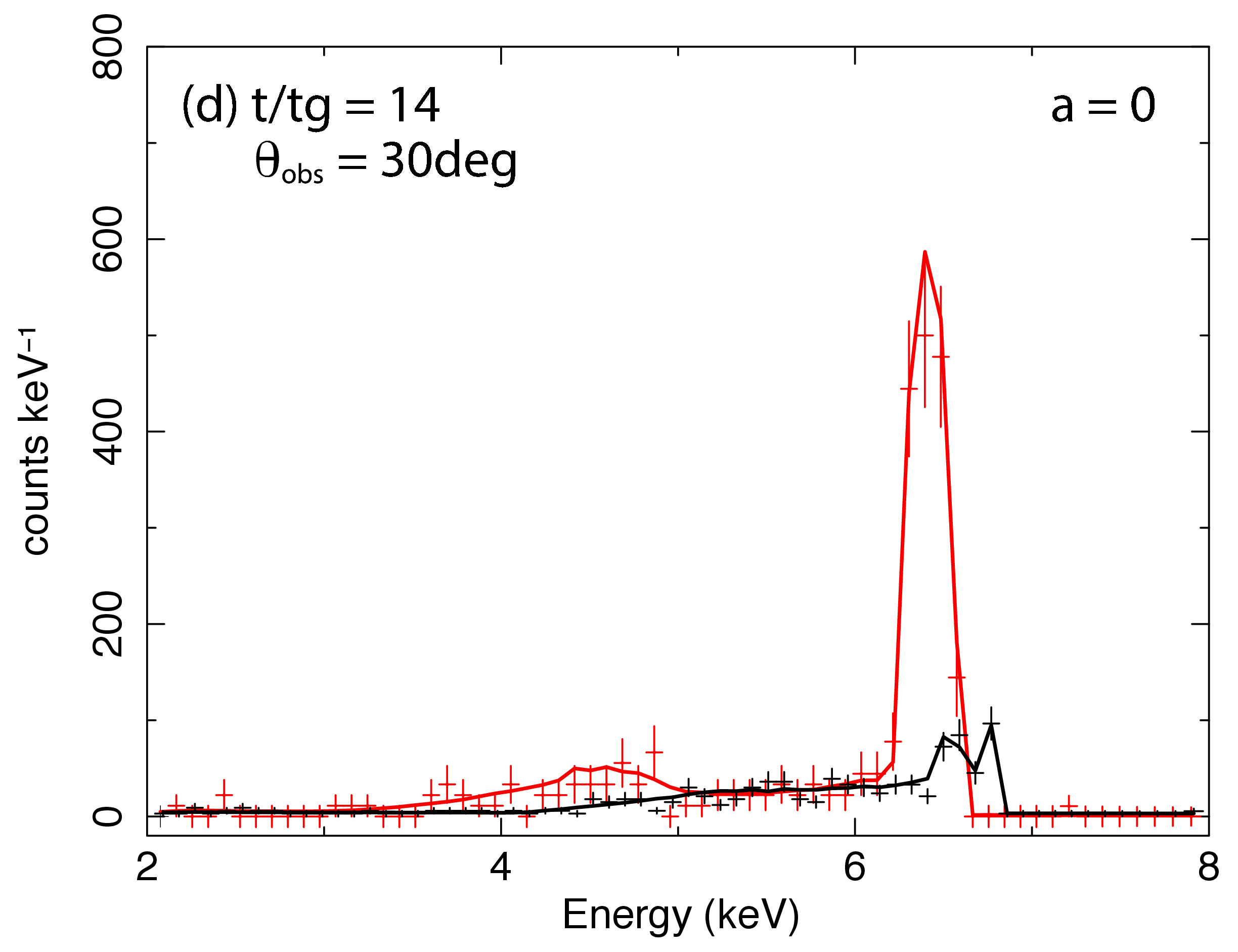} \\
\includegraphics[trim=0in 0in 0in
0in,keepaspectratio=false,width=3.0in,angle=-0,clip=false]{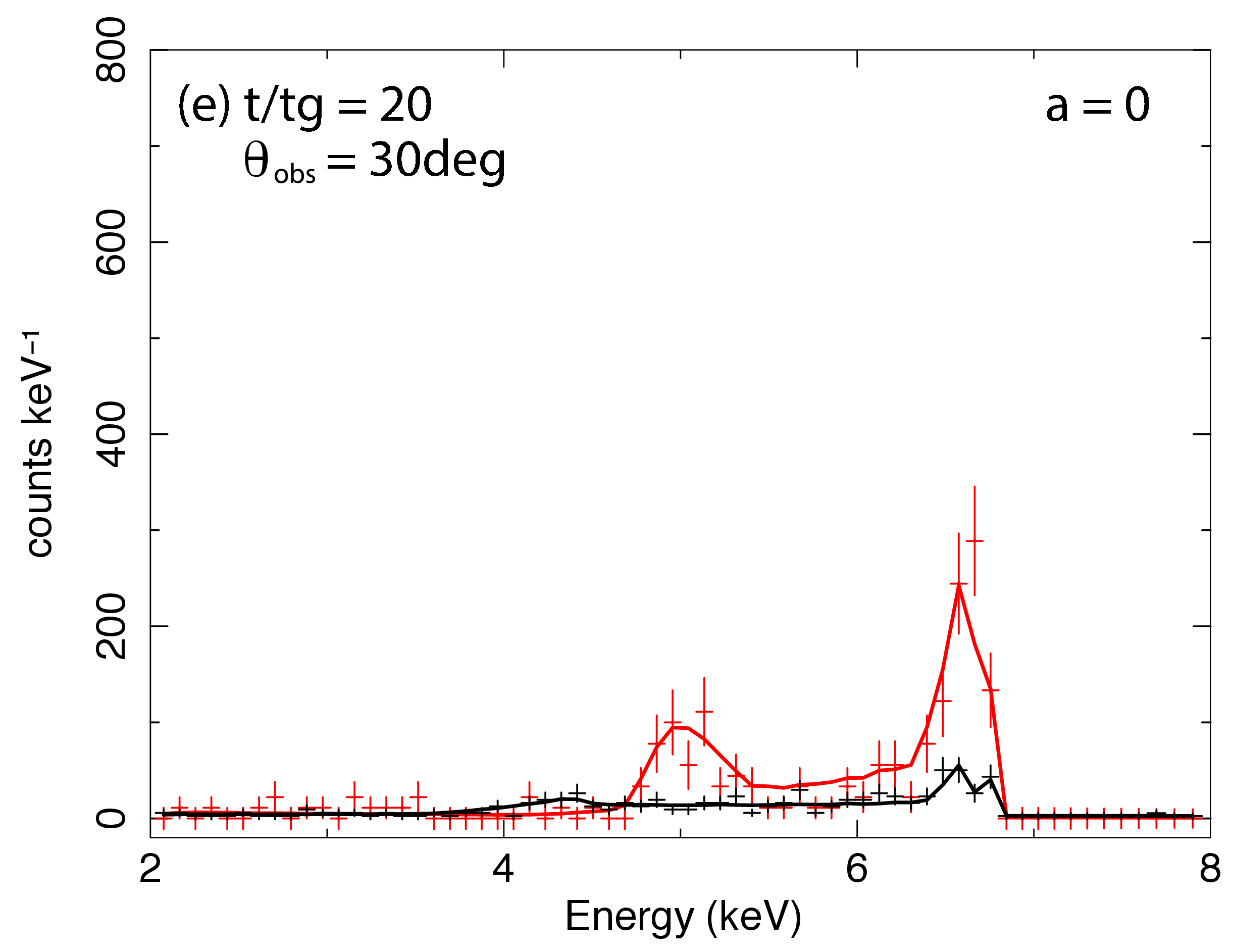}\includegraphics[trim=0in 0in 0in
0in,keepaspectratio=false,width=3.0in,angle=-0,clip=false]{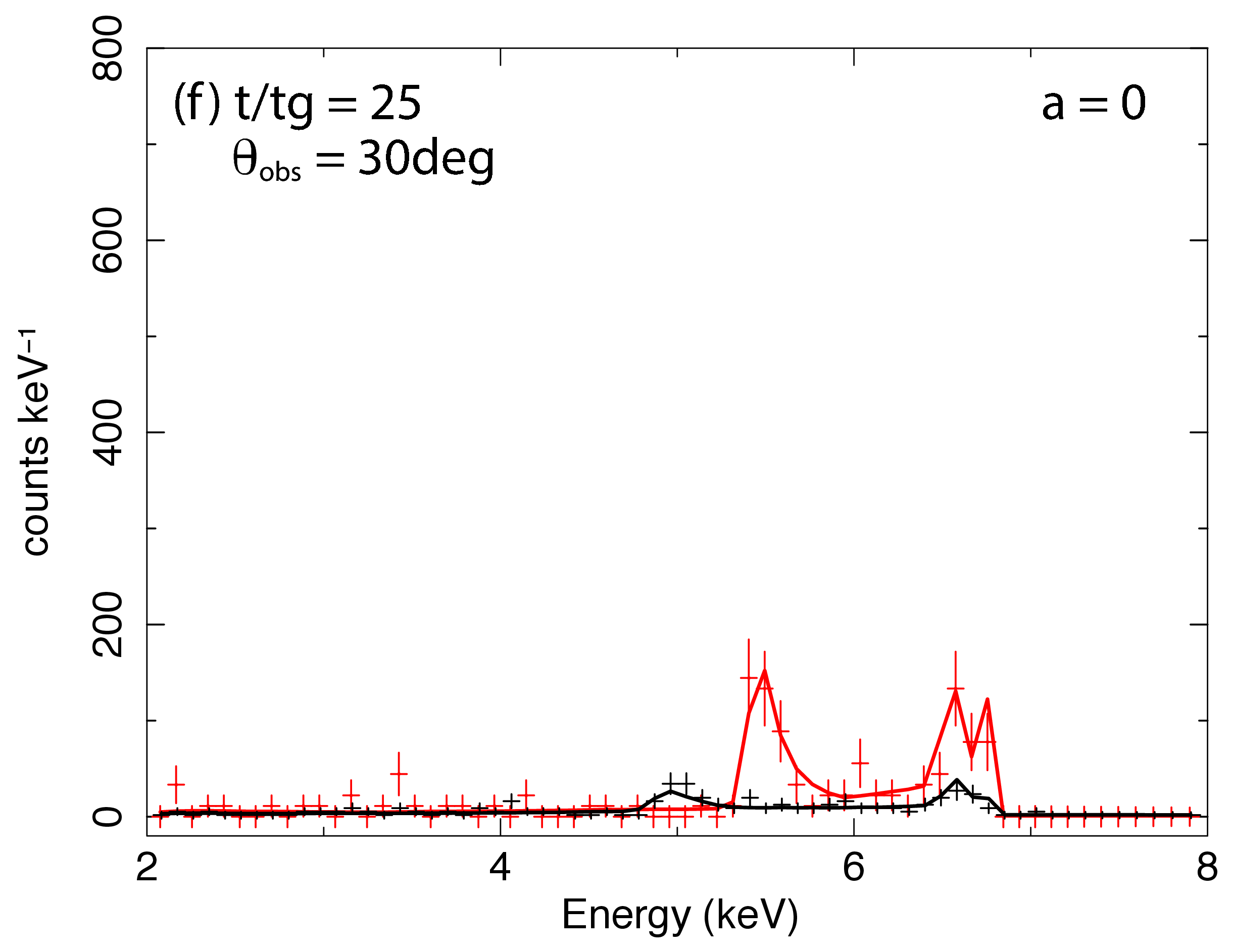}
\end{array}$
\end{center}
\caption{Same as Figure~\ref{fig:spec_a099} but around a \sw BH ($a=0$). The line flux in case (A) is reduced by factor of four for presentation purpose.} \label{fig:spec_a0}
\end{figure}

\clearpage

Qualitatively, the above trend is also the case for a \sw BH, as shown in {\bf Figure~\ref{fig:spec_a0}}.
The simulated spectra in both cases (A) and (B)  exhibit a narrow double-peak with the spectrum in case (A) being systematically more redshifted as similarly found in Kerr case (see {\bf Fig.~3}). It is noted that the extension of the red tail around a \sw BH is much less dramatic because of the larger size of the horizon (i.e. $2r_g$ instead of $\sim r_g$) where gravitational redshift in the innermost region is relatively weaker. Furthermore, the irradiating wavefront from the flare in this case propagates somewhat radially away from the \sw BH in the absence of frame-dragging, which effectively prevents ionizing X-ray photons from being stagnated in the accretion inner region. As the wavefront continues to sweep through a distant portion of the accretion flow, the red peak(s) are seen to progressively shift towards 6.4 keV energy as earlier discussed in Kerr case in {\bf Figure~\ref{fig:spec_a099}}.   

\begin{figure}[ht]
\begin{center}$
\begin{array}{cl}
\includegraphics[trim=0in 0in 0in
0in,keepaspectratio=false,width=3.0in,angle=-0,clip=false]{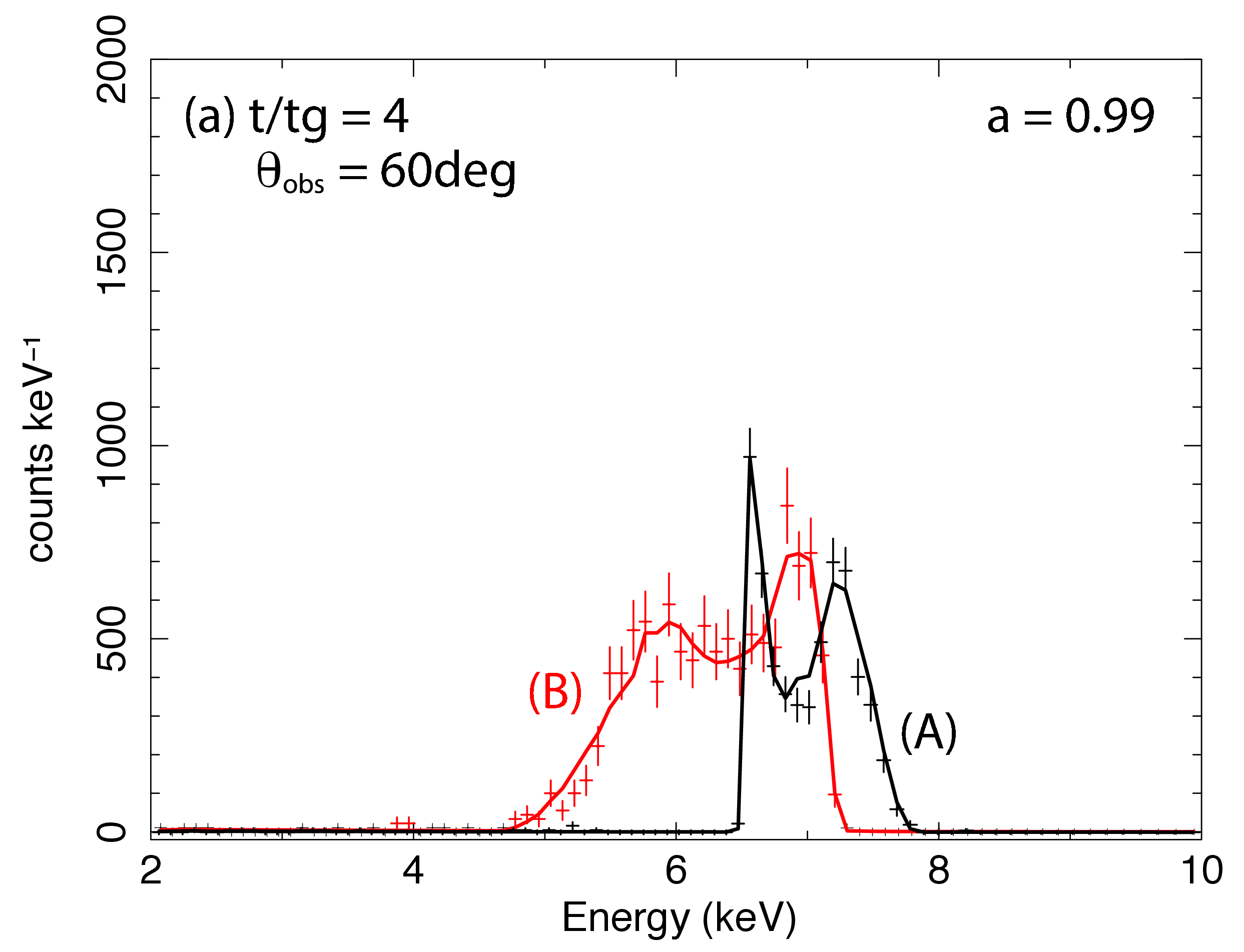}\includegraphics[trim=0in 0in 0in
0in,keepaspectratio=false,width=3.0in,angle=-0,clip=false]{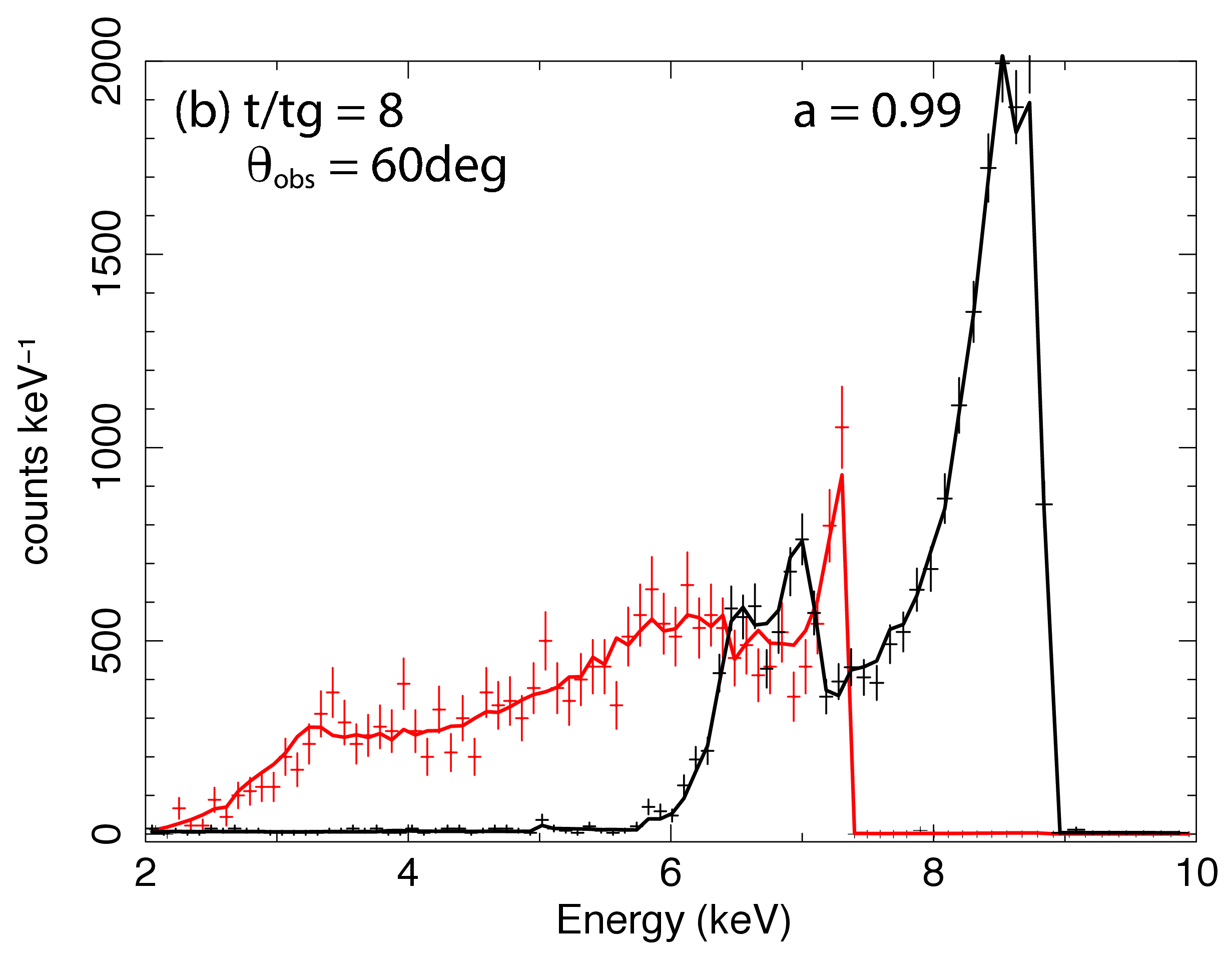} \\
\includegraphics[trim=0in 0in 0in
0in,keepaspectratio=false,width=3.0in,angle=-0,clip=false]{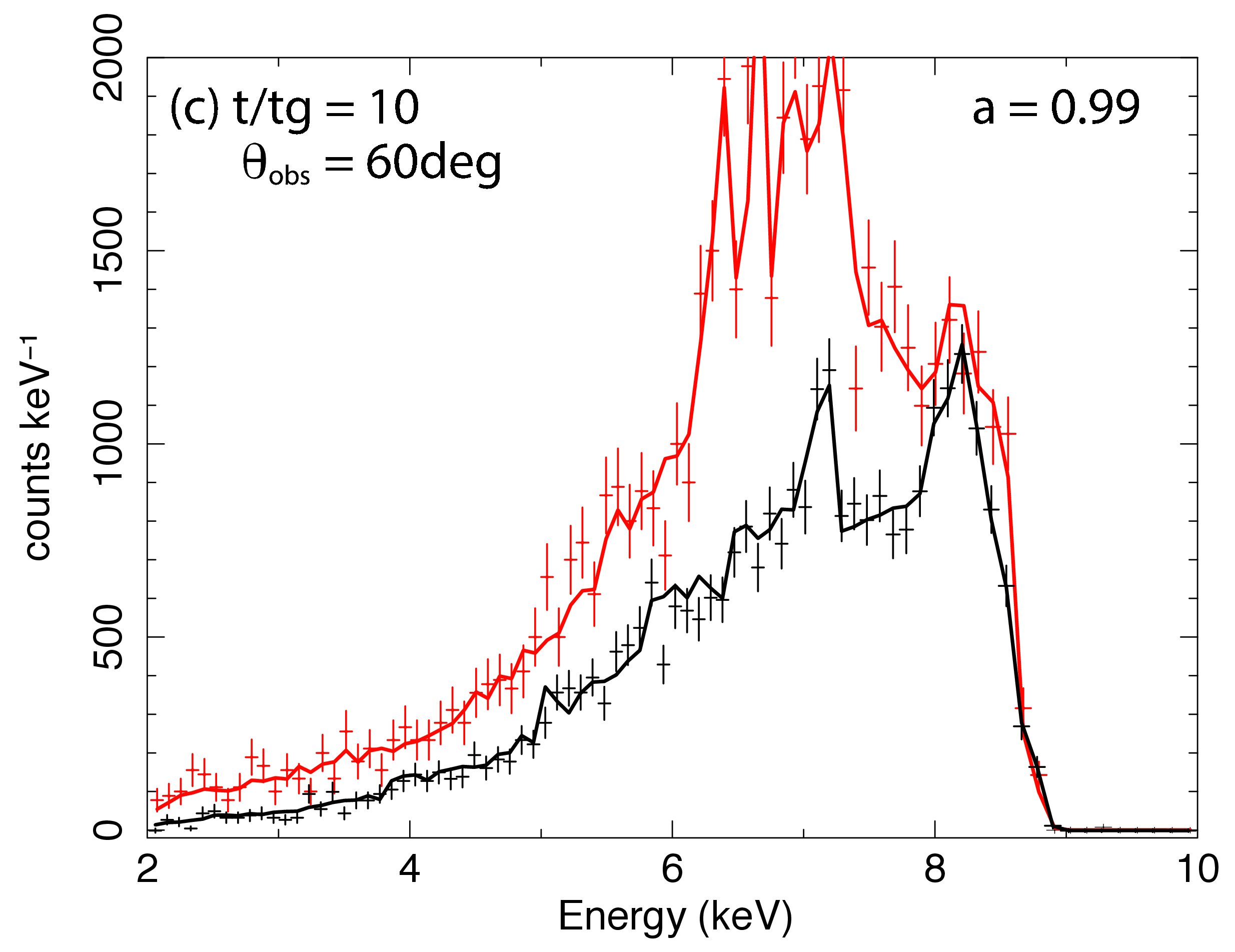}\includegraphics[trim=0in 0in 0in
0in,keepaspectratio=false,width=3.0in,angle=-0,clip=false]{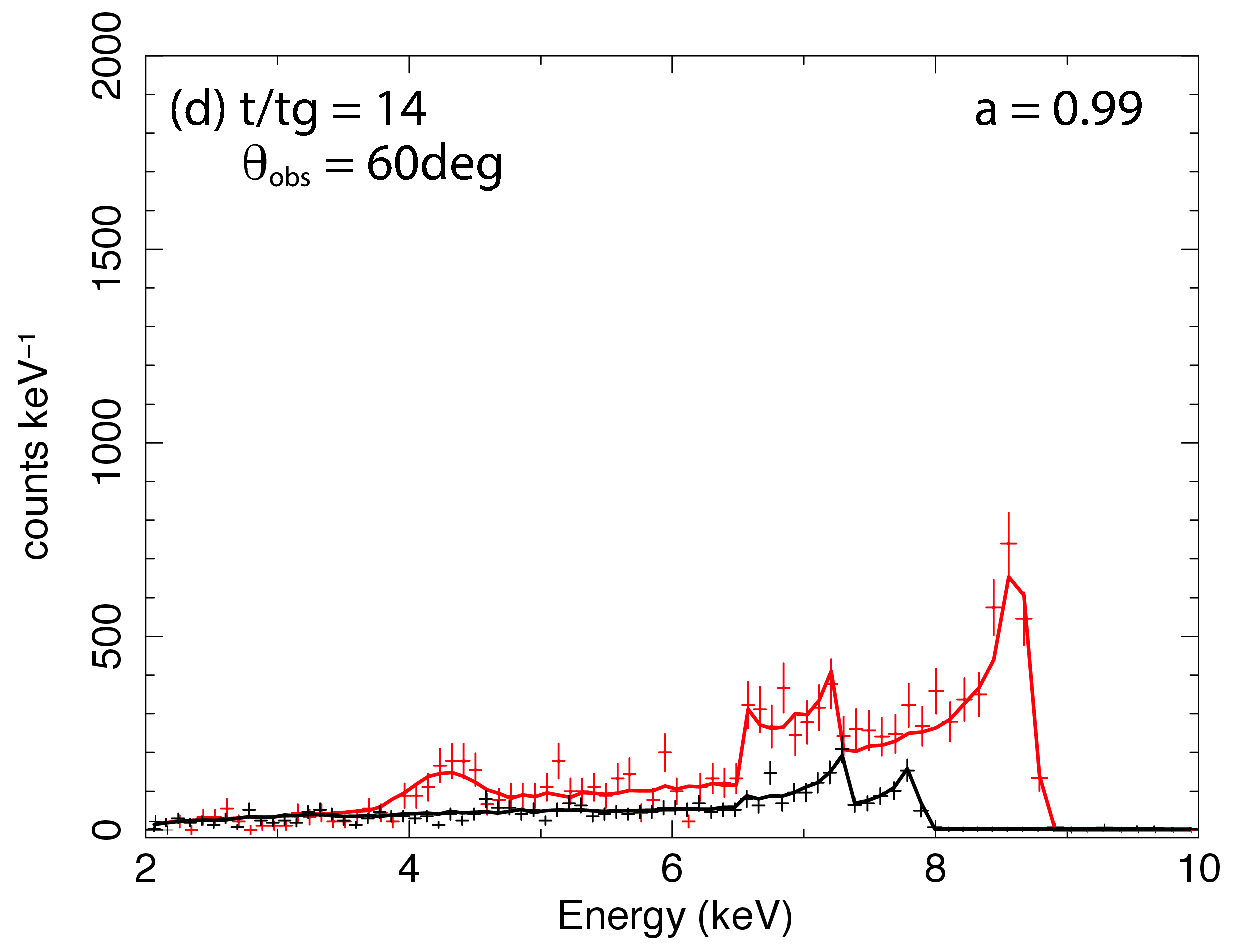} \\
\includegraphics[trim=0in 0in 0in
0in,keepaspectratio=false,width=3.0in,angle=-0,clip=false]{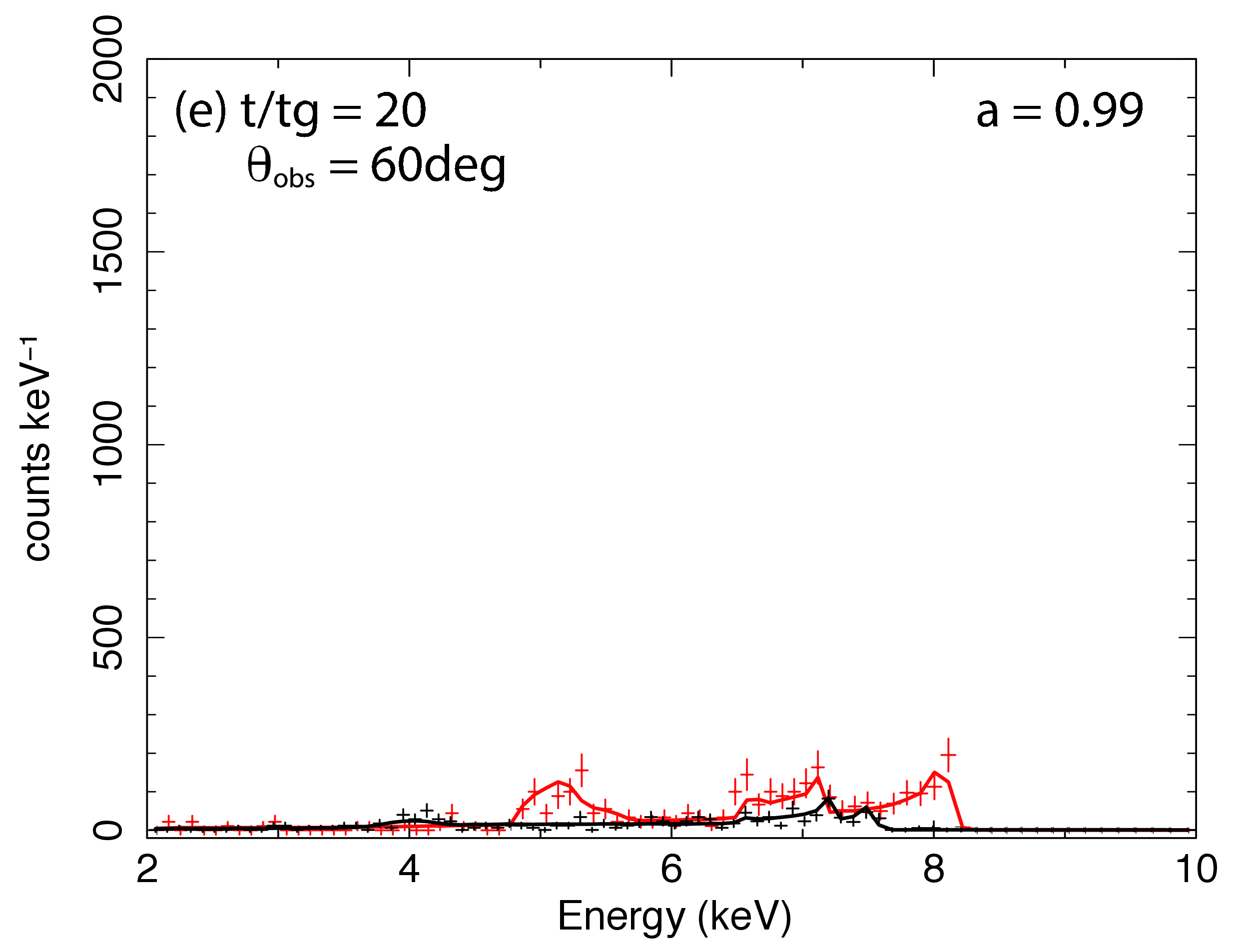}\includegraphics[trim=0in 0in 0in
0in,keepaspectratio=false,width=3.0in,angle=-0,clip=false]{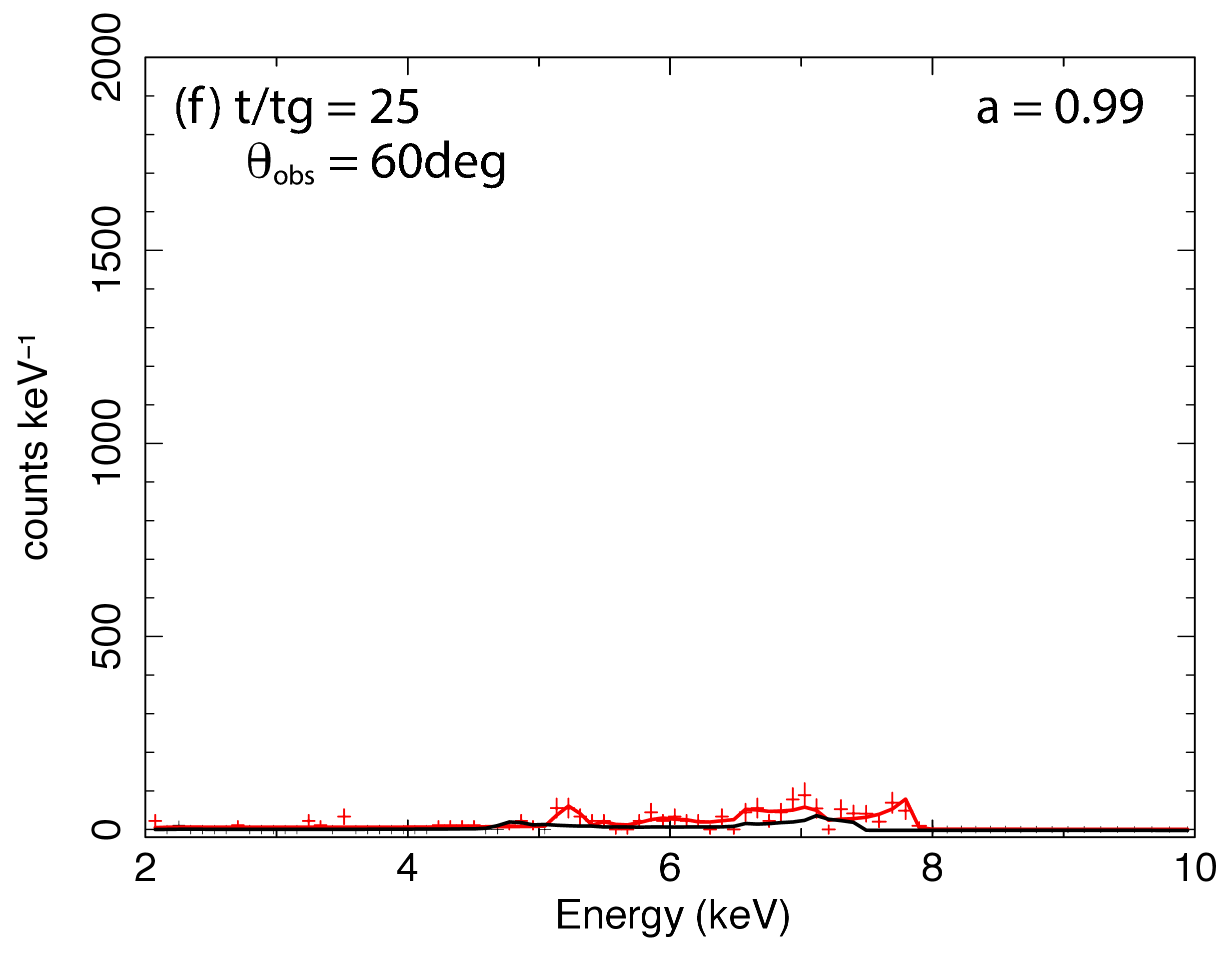}
\end{array}$
\end{center}
\caption{Same as Figure~\ref{fig:spec_a099} but $\theta_{\rm obs}=60\deg$. } \label{fig:spec_a099_60deg}
\end{figure}

In an attempt to further search for a coherent signature of spectral variability due to the X-ray flaring echo, we also investigate the dependence of inclination angle $\theta$ by choosing $\theta=60\deg$. In this case, the line profile is expected to be more broadened especially for the blue peak and its cutoff energy, which is predominantly governed by the viewing angle \citep[e.g.][]{Fabian89,Laor91,Kojima91}. In {\bf Figure~\ref{fig:spec_a099_60deg}} we show the calculated line spectra for $\theta=60\deg$ and $a=0.99$ and it is clearly seen that the line width in general is larger than that for $\theta=30\deg$ case because of the extended blue tail at $E \gsim 8$ keV. The double peak feature appears to be more pronounced as expected. It is noted, uniquely in this high inclination case, that the blue tail cutoff energy is initially $\sim 8$ keV at $t/t_g \sim 4$ and rapidly increases up to $\sim 9$ keV  at $t/t_g \sim 10$, while the red tail becomes gradually extended down to lower energy ($E \lsim 4$ keV) as the X-ray echo sweeps through the approaching side of the gas and the innermost accretion region at the same time. As the flare passes through the horizon at $t/t_g \gsim 14$, we see that the red peak begins to shift back towards $6.4$ keV and the corresponding blue peak energy turns to decrease from $\sim 9$ keV with its flux being gradually reduced due to decreasing gas density with increasing radius, which is a similar behavior found for $\theta=30\deg$ as shown in {\bf Figure~\ref{fig:spec_a099}}. Therefore, we confirm that shifting of the red and blue peaks with time, as found in the other cases ( {\bf Figs.~\ref{fig:spec_a099} and  \ref{fig:spec_a0}}), is indeed a characteristic spectral signature manifesting the X-ray echo through the accreting gas.

\clearpage

\begin{figure}[t]
\begin{center}
\includegraphics[trim=0in 0in 0in
0in,keepaspectratio=false,width=3in,angle=-0,clip=false]{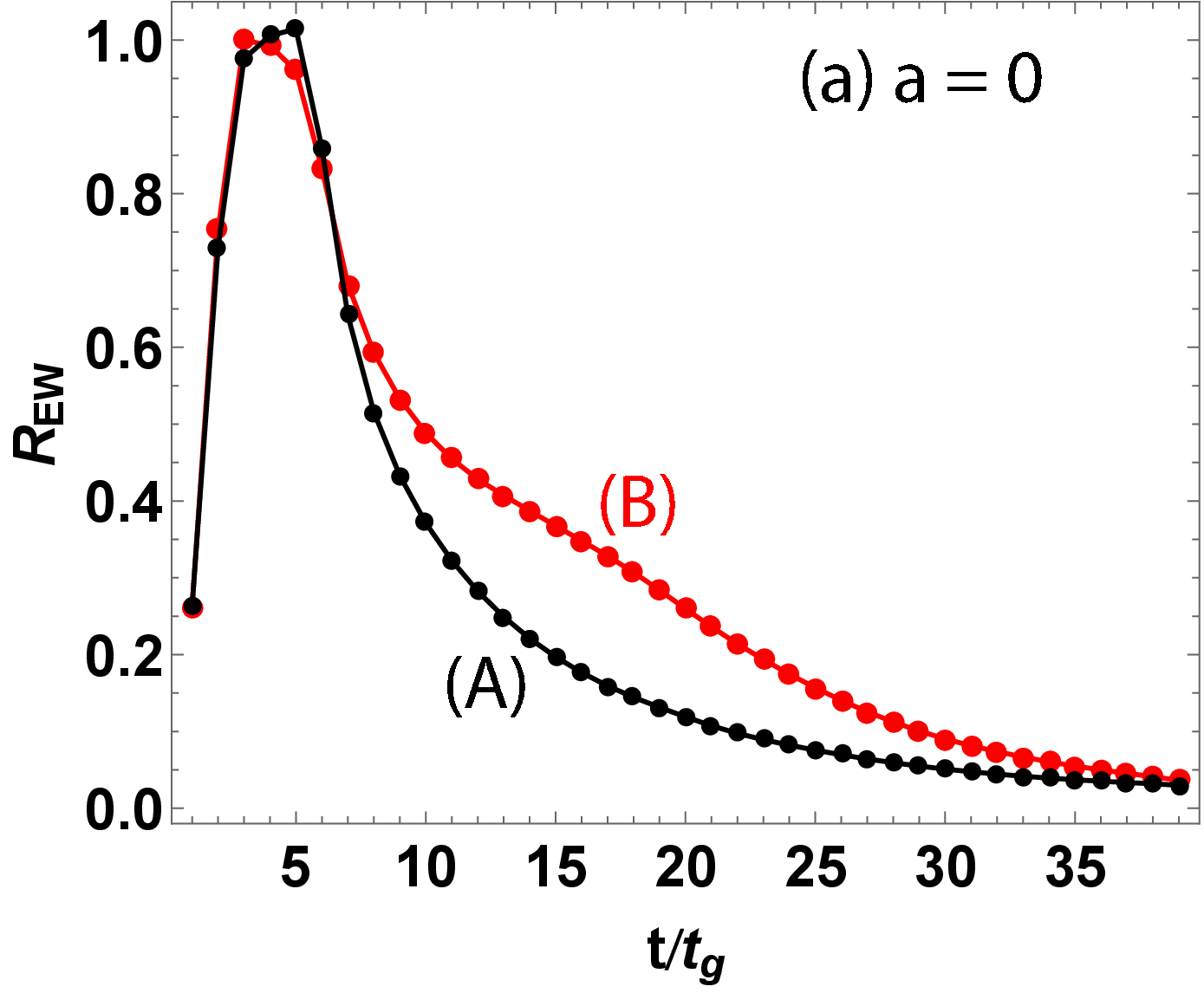}
\includegraphics[trim=0in 0in 0in
0in,keepaspectratio=false,width=3in,angle=-0,clip=false]{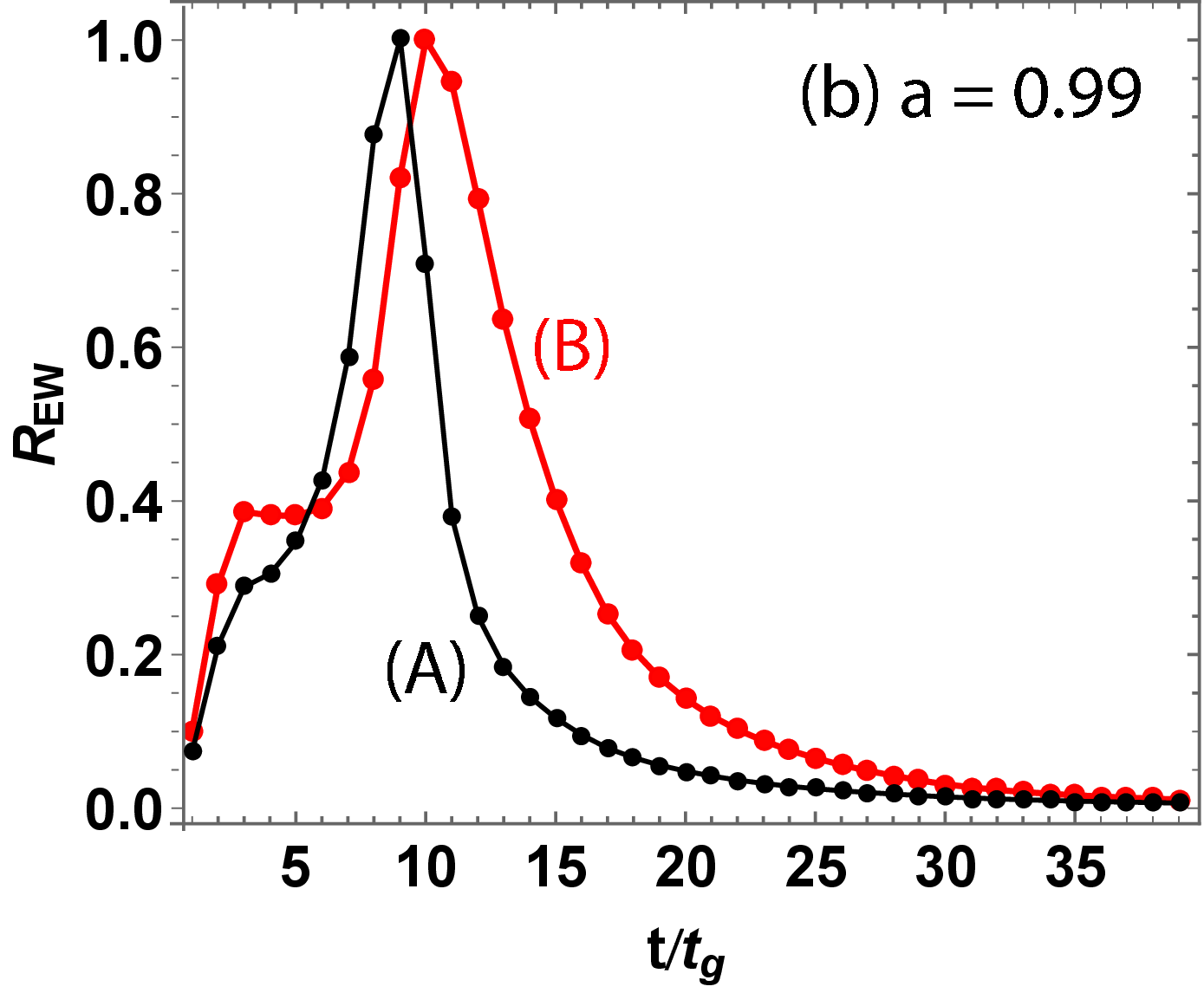}
\end{center}
\caption{Normalized equivalent width (EW), $R_{EW} \equiv EW/EW_{\rm max}$, of the iron emission spectra corresponding to {\bf Figures~\ref{fig:spec_a099} and \ref{fig:spec_a0}} for case (A) in black and case (B) in red. } \label{fig:EW}
\end{figure}

To better evaluate the line strength, the equivalent width (EW) is calculated in {\bf {\bf Figure~\ref{fig:EW}}} corresponding to {\bf Figures~\ref{fig:spec_a099} and \ref{fig:spec_a0}}. It is normalized to the maximum EW in each case to compare the characteristic variability. The line flux at the onset of the flare is low as expected due to the lack of illumination at the initial time. As more area starts being irradiated by the flare ($t/t_g \sim 4$),  there is a rapid increase in the line strength. 
The overall qualitative variability of the predicted EW is almost independent of both BH spin and the position of the flare; i.e. a quick rise and gradual fall.  
There can be seen, however, a slight time delay in the occurrence of the EW peak between \sw case and Kerr case primarily due to the way the X-ray echo propagates. For a \sw BH, the illumination of the flare onto the gas reaches maximum right after the duration of the flare (i.e. at $t/t_g \sim 4$), and the ionizing photon flux monotonically declines with time as the wavefront recedes almost radially away from the \sw BH with the lack of frame-dragging.  
On the other hand, for a Kerr BH, the EW peak occurs at later time (i.e. $t/t_g \sim 10$) long after the duration of the flare because the illuminating X-ray photons in the wavefront are forced to azimuthally rotate around the Kerr BH due to the frame-dragging, which helps enhance the line strength as earlier shown in {\bf Figure~\ref{fig:spec_a099}b-c}. 
We have calculated the EW for $\theta=60\deg$ case and the overall EW variation is found to be qualitatively similar to that for $\theta=30\deg$ case except that the absolute EW values are systematically larger for $\theta=60\deg$.

\section{Discussion \& Conclusions}

In this work we have presented a GRHD model to simulate dynamical iron line profiles through reverberation signatures originating from a short-lived, point-like X-ray flare just above an accretion flow that is relevant for radio-quiet Seyfert 1 AGNs. By numerically calculating GRHD accretion solutions instead of the conventional Keplerian flows, we computed relativistically broadened line spectra and modeled how the line shape would respond to the flare as irradiation is propagating through the gas as a function of time. Theoretical line spectra are then folded into {\it Athena}/X-IFU response matrices, as an example, to create a time-evolution of the simulated spectra for $1$ ks integration. We considered two flare locations, $\phi= \pi$ and $\pi/2$  around a Kerr and a \sw BH of mass $M=10^8 \Msun$, for comparison. The flare is assumed to last for $T_f = 2 t_g \sim 1$ ks in our calculations.

We note that a characteristic red wing is present in all line profiles ({\bf Figures~\ref{fig:spec_a099} and \ref{fig:spec_a0}}). This is a result of gravitational redshift when the flare is approaching/reaches the BH event horizon. This red wing is found to be even broader for flares occurring around a Kerr BH because more photons tend to spend longer time orbiting around a Kerr BH inside the ergosphere due to frame-dragging, a unique characteristics of photon trajectories around a rotating BH, which is absent around a \sw BH. 
Once the wavefront of the X-ray flare has curved around the BH and begins to propagate away from the BH, the red peak (sometimes appearing as multiple red bumps)  
begins to move towards the rest-frame energy at 6.4 keV.
The line spectra around a Kerr BH are consistently broader and systematically more redshifted than those around a \sw case, as expected, for both flare positions. For a Kerr BH, the most extreme redshift occurs within the ergosphere, but the lack of the ergosphere around a \sw BH yields less extreme redshifted photons. To study the line strength, we calculated the expected line EW and it is found to vary in time in a qualitatively similar manner independent of BH spin and flare position. 
For Kerr BH, we note that the line EW becomes maximum, not at the late stage of the onset of the flare, but long after the end of flaring when the ionizing photons from the wavefront are wrapped around and stagnated in the innermost region outside the horizon due to frame-dragging and stronger curvature (see {\bf Fig.~\ref{fig:redshift}}) where the gas density is also higher (see {\bf Fig.~\ref{fig:accretion}}). These factors combined together produce the delayed EW peak in Kerr case ({\bf Fig.~\ref{fig:EW}}).

Many of the previous line reverberation models have assumed either the standard Keplerian accretion flows (e.g. R99; R00; YR00), a non-Keplerian accretion under the pseudo-Newtonian potential \citep[e.g.][]{Armitage03,Chartas17} or restricted to a \sw\ BH \citep[e.g.][]{Schnittman13}. Many theoretical calculations have also adopted a lamp-post model where a point-like X-ray source is postulated to reside at some small height $h$ above the disk near the symmetry axis (e.g. R99; R00; YR00; \citealt[][]{Miniutti03,Miniutti04,Garcia14,Kinch16,Kinch19}).  In the current model, however, we have utilized GRHD accretion for which a flaring event is assumed to occur right above the accreting gas. Such a low-latitude explosion thus progressively illuminates the GRHD gas causing iron fluorescence from the region being illuminated. Therefore, the situation considered here in this work is more appropriate for an energetic flare triggered by a bursting activity on the surface of accretion perhaps somewhat similar to   dwarf novae \citep[e.g.][]{PattersonRaymond85,Mukai17}, accretion shocks \citep[e.g.][]{FukKaz07b,Le16,Le18} or propagation of nuclear burning on accreting neutron stars \cite[e.g.][]{Spitkovsky02}.
The reverberation from an off-equatorial flare geometry of finite height $h$, while interesting and important, is thus beyond the scope of this work.

Among the important factors to determine the line spectrum is ionization state of gas under X-ray irradiation and this can be conveniently quantified as ionization parameter defined as
\begin{eqnarray}
\xi \equiv  \frac{4\pi F_X(r,\phi)}{n(r,\phi)} \ ,
\end{eqnarray} 
where $n(r,\phi)$ is the gas density in accretion at a gas position ($r,\phi$) and $F_X(r,\phi)$ is 
irradiating X-ray flux of the source (defined over some fixed energy band) at position ($r,\phi$). In the present calculations, we simply assume that all the gas is at most only weakly ionized such that cold line photons of 6.4 keV are produced (i.e. Fe K$\alpha$ lines). Given the gas density distribution calculated in the model (see {\bf Fig.~\ref{fig:accretion}}), we should in principle be able to evaluate $\xi(r,\phi)$ for a given flux $F_X$ in an attempt to further assess the line photon energy; e.g. 6.4 keV (neutral/weakly-ionized iron), 6.7 keV (He-like iron) and 6.9 keV (H-like iron), for example, as was done in the past (e.g. \citealt{R97}; R99; R00). 
While the consideration of ionization state would be realistically important and relevant especially for spectral fitting against real data, such a detailed modeling would further introduce more free parameters (e.g. mass-accretion rate $\dot{m}$ for the gas density $n(r,\phi)$ and radiative efficiency $\eta_X$ for the flux $F_X(r,\phi)$). Since our primary focus of this work is placed on the study of a canonical temporal behavior of the broad line profile, we will leave this extension for future work.
It is also noted that a more physically comprehensive study would require a fully numerical 3D global simulation. For example, more  sophisticated GRMHD simulations have been  performed in recent years by self-consistently integrating a Monte Carlo relativistic radiation transport (including ionization state) into the dynamics of fully 3D GRMHD accretion flows to study Fe K$\alpha$ profiles with little phenomenological treatment \citep[e.g.][]{Kinch16,Kinch19}.

In a more realistic picture, a flare source is likely to be in time-dependent motion perhaps coupled to the dynamics of accretion. In that case, the illumination pattern would be a function of time and the incident X-ray flux could be sufficiently anisotropic due to special relativistic beaming effect \citep[e.g.][]{Ruszkowski00,FukKaz07a,WF12,Dauser13}, which could modify the resulting production of fluorescent line photons. For example, if a point-like flare corotates with accreting plasma in the innermost region (e.g. $v^\phi \sim 0.5c$; see {\bf Fig.~1}), the characteristic arc-length displacement of the flare  can be estimated as
$\ell_{\rm flare} \sim V_{\rm flare} \Delta t_{\rm int} \sim v^{\phi} \Delta t_{\rm int} \sim 0.5c \times 10^3 \textmd{[sec]} \sim  R_S/2$ during 1ks integration time for the assumed $M=10^8 \Msun$ BH as considered in this work. In terms of the central illumination point, this is not a terribly large displacement and we feel it justified. 
The flare kinematics, on the other hand, could be important for anisotropically beaming X-ray incident onto the gas  in the gas rest-frame. As the spatial distribution of weakly/neutral and highly ionized Fe atom (discussed earlier) would depend on the illumination pattern, the time-dependent motion of the flare would be important to calculate the ionization parameter $\xi$ of the gas, which could alter the line profiles. While this is beyond the scope of the current work, this is an additional factor to be incorporated for more realistic simulations.

Understanding that one of the defining physical parameters to describe BHs is spin (i.e. intrinsic angular momentum), we have considered both \sw and Kerr BHs in this work.
Constraining BH spin is indeed one of the fundamentally important outstanding questions in AGN/XRB physics. However, the effect of BH spin is very elusive observationally and can be meaningfully extracted only from the innermost accretion region near the ergosphere. For example, X-ray spectral observations so far have attempted to utilize a couple of methods that could potentially provide a way to estimate BH spin; e.g. thermal disk continuum method \citep[e.g.][]{Li05,Shafee06, McClintock14} and relativistic Fe line profiles as a part of disk reflection component \citep[e.g.][]{BR06,Garcia14,Miller07}. However, it has been known that the two methods, independently used for a given object, often derive statistically different BH spin values. 
In addition, the uncertainties in spin measurements derived from individual methods themselves are typically very large ranging from, for example, intermediate to high values, especially for AGNs \citep[e.g.][]{Reynolds14,Vasudevan16}. BH accretion disk spectrum is also known to be degenerate between spin parameter and disk truncation radius \citep[e.g.][]{Dauser13} which makes it more challenging to draw a definitive conclusion as well.  
%
%
Being aware of  those dedicated but enigmatic BH spin measurement, the primary goal of our current work is not focused on solving such a challenging task, but more motivated towards extracting a systematic  spectral signature characteristic to  an ``X-ray echo" due to flares in AGNs. Combined with expected parameter degeneracies among flare location, flare strength, inclination and perhaps BH spin, a more thorough reverberation calculation would need to be conducted in a more comprehensive theoretical framework by constructing a complete sample of template line profile calculations as a function of flare property and its time-evolution, which is beyond the scope of this work.


The present study is focused primarily on a qualitative dynamical nature of the expected line spectra in response to the propagating X-ray flare for different flare position $(r_f,\phi_f)$ with different BH spin ($a=0$ and $0.99$). Obviously, the modeled spectrum would be more time-averaged for a longer exposure, thus reducing finer spectral signatures. For a shorter exposure, on the other hand, more detailed spectral features would remain preserved in theory, while poorer photon statistics would probably prevent them from being robustly detected at a statistically significant level. Hence, the resulting spectrum will depend sensitively on the combination of the allowed exposure, the intrinsic brightness of the source, the detector's collecting area and so on, which is beyond the scope of the current work. In this work, we have instead shown that a certain temporal characteristic of the spectral shape can be identifiable and possibly traceable as a function of time; i.e. well-defined red bumps/peaks are seen to initially shift towards 
lower energy as the ionizing wavefront approaches the horizon and later shift towards the rest-frame energy (at 6.4 keV) after the X-ray echo passes through the horizon and continue to sweep through the gas away from the BH. A similar feature is in fact found in R99, R00 and  YR00 in which a red bump, characteristic to a Kerr BH, shifts towards lower energies over time as the flare approaches the horizon in consistence with this work.


Our simulations suggest that future X-ray missions of high-throughput with large collecting area  will be capable of probing in details the innermost regions of accreting gas responsible for producing the complex iron line spectra with an unprecedented energy resolution using microcalorimeters such as {\it Athena}/X-IFU. High-quality spectra obtained with the upcoming observatories thus may help constrain a putative geometry of the short X-ray flares and eventually BH spin. 


\acknowledgments 

We are grateful to the anonymous referee for their constructive comments and questions. KF appreciates to Truong Lee for his helpful remarks about the initial manuscript and also Demos Kazanas and Peter Becker for their illuminating remarks. This work was supported in part by the Department of Physics \& Astronomy at James Madison University through Jeffery E. Tickle Scholarship awarded by JMU College of Science and Math.


\begin{thebibliography}{999}

\bibitem[Abramowicz et al.(1995)]{Abramowicz95} Abramowicz, M. A., Chen, X., Kato, S., Lasota, J. -P., \& Regev, O. 1995, \apj, 438, L37

\bibitem[Abramowicz et al.(1997)]{Abramowicz97} Abramowicz, M. A., Lanza, A., \& Percival, M. J. 1997, \apj, 479, 179

%

\bibitem[Ar\'{e}valo et al.(2005)]{Arevalo05} Ar\'{e}valo, P., Papadakis, I., Kuhlbrodt, B.,\& Brinkmann, W. 2005, A \& A, 430, 435

\bibitem[Armitage \& Reynolds(2003)]{Armitage03} Armitage, P. J. \& Reynolds, C. S. 2003, \mnras, 341, 1041


\bibitem[Bardeen et al.(1972)]{Bardeen72} Bardeen, J. M., Press, W. H., \& Teukolsky, S. A. 1972, \apj, 178, 347
\bibitem[Beckwith et al.(2008)]{Beckwith08} Beckwith, Kris; Hawley, John F.; Krolik, Julian H.
\bibitem[Blandford \& Begelman(1999)]{BB99} Blandford, R. D., \& Begelman, M. C. 1999, \mnras, 303, L1 

\bibitem[Blandford \& McKee(1982)]{BlandfordMcKee82}Blandford, R. D., \& McKee, C. F. 1982, \apj, 255, 419

\bibitem[Boller et al.(2003)]{Boller03} Boller, T., et al. 2003, A\&A, 411, 63

\bibitem[Brenneman \& Reynolds(2006)]{BR06} Brenneman, L. W., \& Reynolds, C. S. 2006, \apj, 652, 1028

\bibitem[Cackett et al.(2008)]{Cackett08} Cackett, E. M., et al. 2008, \apj, 674, 415

\bibitem[\v{C}ade\v{z} et al.(1998)]{Cadez98} \v{C}ade\v{z}, A., Fanton, C., \& Calvani, M. 1998, New Astronomy, 3, 647

\bibitem[Chakrabarti(1990)]{Chakrabarti90} Chakrabarti, S. K. 1990, Theory of Transonic Astrophysical Flows, World Scientific, Singapore

\bibitem[Chackrabarti(1996)]{Chakrabarti96} Chakrabarti, S. K. 1996, \apj, 464, 664


\bibitem[Chartas et al.(2017)]{Chartas17} Chartas, G., Krawczynski, H., Zalesky, L., Kochanek, C. S., Dai, X., Morgan, C. W., \& Mosquera, A. 2017, \apj, 837, 26

%
%

\bibitem[Dauser et al.(2013)]{Dauser13} Dauser, T. et al. 2013, \mnras, 430, 1694


\bibitem[Fabian et al.(2000)]{Fabian00} Fabian, A. C., Iwasawa, K., Reynolds, C. S., \& Young, A. J. 2000, PASP, 112, 1145

\bibitem[Fabian et al.(1989)]{Fabian89} Fabian, A.C., Rees, M. J., Stella, L., \& White, N. E. 1989, \mnras, 238, 729

\bibitem[Fanton et al.(1997)]{Fanton97} Fanton, C., Calvani, M., de Felice, F., \& Cadez, A. 1997, \pasj, 49, 159

\bibitem[Fukumura \& Tsuruta(2004)]{FT04} Fukumura, K. \& Tsuruta, S. 2004, ApJ, 611, 964

\bibitem[Fukumura \& Kazanas(2007a)]{FukKaz07a} Fukumura, K., \& Kazanas, D. 2007a, \apj, 664, 14
\bibitem[Fukumura \& Kazanas(2007b)]{FukKaz07b} Fukumura, K., \& Kazanas, D. 2007b, \apj, 669, 85



\bibitem[Garc\'{i}a et al.(2014)]{Garcia14} Garc\'{i}a, J., Dauser, T., Lohfink, A., Kallman, T. R., Steiner, J. F., McClintock, J. E., Brenneman, L., Wilms, J., Eikmann, W., Reynolds, C. S. \& Tombesi, F. 2014, ApJ, 782, 76


\bibitem[Hawley \& Krolik(2001)]{HawleyKrolik01} Hawley, J. F. \& Krolik, J. H. 2001, ApJ, 548, 348

\bibitem[Ishida et al.(2009)]{Ishida09} Ishida, M., Okada, S., Hayashi, T., Nakamura, R., Terada, Y., Mukai, K., \& Hamaguchi, K. 2009, \pasj, 61, S77

%
%
%
%
%


\bibitem[Kinch et al.(2016)]{Kinch16} Kinch, B. E., Schnittman, J. D., Kallman, T. R. \& Krolik, J. 2016, ApJ, 826, 52  
\bibitem[Kinch et al.(2019)]{Kinch19} Kinch, B. E., Schnittman, J. D., Kallman, T. R. \& Krolik, J. 2019, ApJ, 873, 71  
\bibitem[Kojima(1991)]{Kojima91} Kojima, Y. 1991, \mnras, 250, 629

\bibitem[Laor(1991)]{Laor91} Laor, A. 1991, \apj, 376, 90

\bibitem[Le et al.(2016)]{Le16} Le, T., Wood, K. S., Wolff, M. T., Becker, P. A., Putney, J. 2016, ApJ, 819, 112
\bibitem[Le et al.(2018)]{Le18} Le, T., Newman, W. \& Edge, B. 2018, MNRAS, 477, 1803 
\bibitem[Lee et al.(1999)]{Lee99} Lee, J. C., Fabian, A. C., Brandt, W. N., Reynolds, C. S., \& Iwasawa, K. 1999, \mnras, 310, 973
\bibitem[Li et al.(2005)]{Li05} Li, L.-X., Zimmerman, E. R., Narayan, R. \& McClintock, J. E. 2005, ApJS, 157, 335

\bibitem[Lu et al.(1997)]{Lu97} Lu, J. -F., Yu, K. N., Yuan, F., \& Young, E. C. M. 1997, A\&A, 321, 665

\bibitem[Lu \& Yuan(1998)]{Lu98} Lu, J.-F. \& Yuan, F. 1998, \mnras, 295, 66

\bibitem[Manmoto(2000)]{Manmoto00} Manmoto, T. 2000, \apj, 534, 734

\bibitem[McClintock et al.(2014)]{McClintock14} McClintock, J. E., Narayan, R., \& Steiner, J. F. 2014, Space Science Reviews, 183, 295

\bibitem[Miller(2007)]{Miller07} Miller, J. M. 2007, ARA\&A, 45, 441

\bibitem[Miniutti et al.(2003)]{Miniutti03} Miniutti, G., Fabian, A. C., Goyder, R., \& Lasenby, A. N. 2003, \mnras, 344, L22

\bibitem[Miniutti \& Fabian(2004)]{Miniutti04} Miniutti, G. \& Fabian, A. C. 2004, \mnras, 349, 1435



\bibitem[Mitsuda et al.(1984)]{Mitsuda84} Mitsuda, K., Inoue, H., Koyama, K., Makishima, K., Matsuoka, M., Ogawara, Y., Shibazaki, N., Suzuki, K., Tanaka, Y., \& Hirano, T. 1984, \pasj, 36, 741
\bibitem[Mukai(2017)]{Mukai17} Mukai, K. 2017, PASP, 129, 976
\bibitem[Nandra et al.(2007)]{Nandra07} Nandra, K., O'Neill, P. M., George, I. M., \& Reeves, J. N. 2007, \mnras, 382, 194

\bibitem[Narayan \& Yi(1994)]{NY94} Narayan, R., \& Yi, I. 1994, \apj, 428, L13

\bibitem[Narayan et al.(2000)]{Narayan00} Narayan, R., Igumenshchev, I. V., Abramowicz, M. A. 2000, ApJ, 539, 798

\bibitem[Novikov \& Thorne(1973)]{Novikov73} Novikov, I. D., \& Thorne, K. S. 1973, in Black Holes,
ed. C. DeWitt and B. DeWitt, Gordon and Breach, New York


\bibitem[Page \& Thorne(1974)]{Page74} Page, D., \& Thorne, K. S. 1974, \apj, 191, 499
\bibitem[Patterson \& Raymond(1985)]{PattersonRaymond85} Patterson, J.; Raymond, J. C. 1985, ApJ, 292, 535

\bibitem[Quataert \& Gruzinov(2000)]{Quataert00} Quataert, E. \& Gruzinov, A. 2000, ApJ, 539, 809

\bibitem[Reynolds \& Begelman(1997)]{R97} Reynolds, C. S. \& Begelman, M. C. 1997, ApJ, 488, 109
\bibitem[Reynolds et al.(1999)]{R99} Reynolds, C. S.,Young, A. J., Begelman, M. C., \& Fabian, A. C. 1999, \apj, 514, 164 (R99)
\bibitem[Reynolds \& Fabian(2008)]{RF08} Reynolds, C. S., Fabian, A. C. 2008, \apj, 675, 1048
\bibitem[Reynolds(2014)]{Reynolds14} Reynolds, C. S. 2014, Space Science Reviews, 183, 277 
\bibitem[Ruszkowski(2000)]{Ruszkowski00} Ruszkowski, M. 2000, \mnras, 315, 1

%
%

\bibitem[Schnittman et al.(2013)]{Schnittman13} Schnittman, J. D., Krolik, J. H., \& Noble, S. C. 2013, \apj, 769, 156

\bibitem[Shafee et al.(2006)]{Shafee06} Shafee, R., McClintock, J. E., Narayan, R., Davis, S. W., Li, L.-X. \& Remillard, R. A. 2006, ApJ, 636, 113

\bibitem[Shakura \& Sunyaev(1973)]{SS73} Shakura, N. I., \& Sunyaev, R. A. 1973, IAU Symposium, 55, 155

\bibitem[Spitkovsky et al.(2002)]{Spitkovsky02} Spitkovsky, A., Levin, Y. \& Ushomirsky, G. 2002, ApJ, 566, 1018 

\bibitem[Stella(1990)]{Stella90}Stella, L. 1990, Nature, 344,747

%
%
%

\bibitem[Takahashi et al.(2002)]{T02} Takahashi, M., Rilett, D., Fukumura, K. \& Tsuruta, S. 2002, ApJ, 572, 950

\bibitem[Tanaka et al.(1995)]{Tanaka95} Tanaka, Y., Nandra, K., Fabian, A. C., Inoue, H., Otani, C., Dotani, T., Hayashida, K., Iwasawa, K., Kii, T., Kunieda, H., Makino, F., \& Matsuoka, M. 1995, Nature, 375, 659
\bibitem[Vasudevan et al.(2016)]{Vasudevan16} Vasudevan, R. V., Fabian, A. C., Reynolds, C. S., Aird, J., Dauser, T. \& Gallo, L. C. 2016, MNRAS, 458, 2012


\bibitem[Wilkins \& Fabian(2012)]{WF12} Wilkins, D. R., \& Fabian, A. C. 2012, \mnras, 424, 1284


\bibitem[Young \& Reynolds(2000)]{YR00} Young, A. J., \& Reynolds, C. S. 2000, \apj,529,101 (YR00)

\end{thebibliography}
\end{document}